\documentclass[a4paper,UKenglish,cleveref,autoref,thm-restate]{lipics-v2021}
\usepackage{here}
\usepackage{booktabs}
\usepackage{mathtools}
\usepackage{placeins}
\usepackage{xcolor}
\usepackage{tikz}
\usepackage{pgfplots}
\usetikzlibrary{arrows.meta,backgrounds,calc,decorations.pathreplacing,fit,matrix,positioning}
\usepgfplotslibrary{groupplots,statistics}
\pgfplotsset{compat=1.18}

\definecolor{cutRed}{HTML}{BD3B3B}
\definecolor{plotBlue}{HTML}{004488}
\definecolor{plotOrange}{HTML}{BB5566}
\definecolor{plotGreen}{HTML}{DDAA33}
\definecolor{plotPurple}{HTML}{DDAA33}
\definecolor{figInk}{HTML}{303238}
\definecolor{figFill}{HTML}{E2E0D9}
\definecolor{figEmph}{HTML}{004488}
\definecolor{figAcc}{HTML}{BB5566}
\definecolor{figCode}{HTML}{ECECE7}

\newcommand{\code}[1]{\texttt{#1}}
\tikzset{
  tr/.style={anchor=north west,inner xsep=1mm,inner ysep=.35mm,minimum height=3.7mm,
             text width=\dimexpr\linewidth-2mm-1pt\relax,align=left,
             font=\scriptsize\ttfamily,text height=1.6ex,text depth=.45ex},
  trctl/.style={tr,text=black},
  trval/.style={tr,fill=figEmph!25},
  leftbar/.style={append after command={\pgfextra{\draw[figAcc,line width=1.6pt,#1]
    (\tikzlastnode.north west) -- (\tikzlastnode.south west);}}},
  trmem/.style={tr,fill=figAcc!25,leftbar=solid},
  trskip/.style={tr,font=\scriptsize\itshape,text=black},
  trgap/.style={tr,leftbar={dash pattern=on 1.2pt off 1.2pt},font=\scriptsize\itshape,text=black},
  trsrc/.style={fill=figCode,draw=figInk!70,rounded corners=1.5pt,inner sep=1.5mm,
                text width=37mm,align=left,font=\scriptsize\ttfamily,anchor=north west},
  trkey/.style={minimum width=5mm,minimum height=2.6mm,inner sep=0pt}
}
\newcommand{\TrRow}[2]{\par\nointerlineskip\vspace{.3mm}\noindent
  \tikz[baseline]\node[#1]{\detokenize{#2}};}
\newcommand{\TrCtl}[1]{\TrRow{trctl}{#1}}
\newcommand{\TrVal}[1]{\TrRow{trval}{#1}}
\newcommand{\TrMem}[1]{\TrRow{trmem}{#1}}
\newcommand{\TrSkip}[1]{\TrRow{trskip}{#1}}
\newcommand{\TrGap}{\par\nointerlineskip\vspace{.3mm}\noindent
  \tikz[baseline]\node[trgap]{stack write removed};}
\newcommand{\TrSrc}[2]{\par\noindent\hspace*{#1\dimexpr.525em\relax}#2}
\newcommand{\TrSrcHL}[3]{\par\noindent\hspace*{#1\dimexpr.525em\relax}%
  {\setlength{\fboxsep}{.6pt}\colorbox{figEmph!40}{#2}}#3}
\newcommand{\TrSource}[1]{\noindent
  \begin{tikzpicture}[baseline=(src.north)]
  \node[trsrc] (src) {#1};
  \node[anchor=north west,font=\scriptsize,align=left] at ([xshift=6mm]src.north east) {%
    \tikz[baseline=-1.5ex]\node[trkey,fill=figEmph!25]{};\ arithmetic of the highlighted loop body\\[1.2mm]
    \tikz[baseline=-1.5ex]\node[trkey,fill=figAcc!25,leftbar=solid]{};\ stack write left in the trace\\[1.2mm]
    \tikz[baseline=-1.5ex]\node[trkey,leftbar={dash pattern=on 1.2pt off 1.2pt}]{};\ the same write, removed by exposure};
  \end{tikzpicture}\par\medskip}
\newcommand{\TrColumn}[3]{%
  \begin{minipage}[t]{.49\linewidth}\setlength{\parskip}{0pt}%
  \noindent{\small\bfseries #1}\par\smallskip
  #3\par\smallskip\noindent
  {\scriptsize\bfseries\ifnum#2=0 \color{figEmph}no stack writes per iteration%
   \else\color{figAcc}#2 stack writes per iteration\fi}%
  \end{minipage}\hfill}
\newenvironment{TrFigure}{\begin{minipage}{\linewidth}\setlength{\parskip}{0pt}}{\end{minipage}}
\newcommand{\system}{RPyForth}
\newcommand{\rpyforth}{RPyForth}
\newcommand{\gf}{gforth-fast}
\newcommand{\swift}{SwiftForth}
\newcommand{\vfx}{VFX Forth}
\newcommand{\layout}{fixed-width top fragment}
\newif\ifsupplemental
\supplementaltrue
\crefname{supplemental}{Supplemental}{Supplementals}
\Crefname{supplemental}{Supplemental}{Supplementals}
\newcommand{\decode}{\mathop{\mathrm{decode}}}
\newcommand{\project}{\mathop{\mathrm{project}}}
\newcommand{\stackseq}{\mathit{stack}}
\newcommand{\spillseq}{\mathit{spill}}
\newcommand{\winseq}{\mathit{win}}
\newcommand{\wmax}{w_{\mathrm{max}}}
\newcommand{\ccall}{c_{\mathrm{call}}}
\newcommand{\bspill}{b_{\mathrm{spill}}}
\newcommand{\dmax}{d_{\mathrm{max}}}
\newcommand{\ntop}{n_{\mathrm{top}}}
\newcommand{\nframe}{n_{\mathrm{frame}}}

\title{Exposing Call-Shared Stacks to a Meta-Tracing JIT Compiler}
\titlerunning{Exposing Call-Shared Stacks to a Meta-Tracing JIT Compiler}

\nolinenumbers
\hideLIPIcs

\author{Yusuke Izawa}{Tokyo Metropolitan University, Japan \and Heinrich-Heine-Universit\"at D\"usseldorf, Germany}{}{}{}
\author{Kota Hakamada}{Tokyo Metropolitan University, Japan}{}{}{}
\authorrunning{Y. Izawa and K. Hakamada}
\Copyright{Yusuke Izawa and Kota Hakamada}

\category{}
\relatedversion{}
\supplement{}
\funding{}
\acknowledgements{We would like to thank Carl Friedrich Bolz-Tereick for their
  valuable comments on the central idea of this paper.}

\begin{document}
\maketitle
\begin{abstract}
Meta-tracing JIT compilers such as RPython's can optimize accesses to
interpreter state when the state is represented in a form that the optimizer can
track, such as fixed fields in virtualizable interpreter frames. However,
interpreters for stack-based languages such as Forth and Factor use a shared
data stack whose depth and per-call usage are not necessarily known
statically. Keeping this stack in a heap array may leave memory accesses in
compiled traces, while exposing the entire array to the optimizer makes the
amount of exposed state depend on its allocated capacity. Conventional stack
caching avoids some memory accesses but does not necessarily make cached values
visible to a meta-tracing JIT compiler.

We propose the stack fragment layout, which divides a shared data stack into a
fixed-width top fragment and a shared spill array. The top fragment stores stack
values in interpreter fields that a meta-tracing JIT compiler can optimize
within traces, while the spill retains deeper values in memory. This
representation bounds the amount of JIT-visible stack state independently of the
total stack depth, without requiring static stack-effect information.
We provide proof sketches showing that the layout preserves the modeled stack
operations and, under stated assumptions, that the amount of stack state
reconstructed at trace exits is bounded by the fragment width plus constant
control state. We implement the layout in RPyForth, an RPython-based Forth
interpreter, and RPyFactor, an interpreter for a subset of Factor.

Controlled experiments isolate the benefit of exposing the top fragment to the
meta-tracing JIT compiler from that of stack caching alone. Across three
benchmark suites on two x86-64 machines, this optimization yields suite-level
geometric-mean speedups of \(1.44\)--\(2.11\times\) in RPyForth and
\(1.42\)--\(1.56\times\) in RPyFactor.
\end{abstract}

\section{Introduction}
\label{sec:introduction}

Meta-tracing frameworks such as RPython~\cite{bolz09,bolztratt15} and
yk~\cite{yk} generate JIT compilers from interpreters. They trace frequently
executed paths through an interpreter and compile the resulting traces into
machine code. Beyond eliminating interpreter dispatch overhead, meta-tracing
enables optimization of the interpreter's state across operations. In RPython,
two mechanisms are particularly effective at eliminating accesses to interpreter
state: \emph{virtuals}, which represent objects whose allocations can be removed
within a trace~\cite{bolz11}, and \emph{virtualizables}, which allow designated
interpreter frame fields to be optimized across
traces~\cite{rpythonvirtualizables}. Both work particularly well with statically
structured state, such as the fixed set of local variables and operand slots in
a bytecode interpreter's activation frame.

Some interpreters share stack storage across calls while retaining statically
determined storage requirements within individual functions. Lua 5.0 evaluates
call arguments into the caller's next free registers, which become part of the
callee's activation record, and preallocates each record large enough for all of
its function's registers~\cite{lua05}. An in-place WebAssembly interpreter similarly
uses the caller's outgoing arguments as the callee's first locals, overlapping
the caller's operand stack on a contiguous value stack. WebAssembly validation
tracks the abstract operand stack within each function body, providing
stack-height information that the interpreter can reuse~\cite{titzer22}. Factor
translates lexical variables into stack operations and normally requires
stack-effect declarations for word definitions, which are checked by its
compiler~\cite{factor10}. A
stack effect specifies the values a word consumes from and produces on the
stack. In contrast, concatenative languages such as Forth (in its core word
set), PostScript, and Joy do not require declared stack effects or separate
per-call operand frames. Their words and procedures consume operands and produce
results on a shared data or operand stack, while call-control information is
maintained separately~\cite{forth2012,plrm3,joymanual}.

Forth illustrates this shared-stack execution model. A Forth program consists of
\emph{words} that operate on a common data stack. For example, invoking
\code{SQUARE}, defined as \code{: SQUARE DUP * ;}, does not create a separate
data-stack frame. Instead, it consumes a value left by the caller and replaces
it with the result~\cite{forth2012}.

Optimizing such a shared stack poses a challenge for existing stack
representations in RPython. When the stack is stored in an ordinary heap array,
accesses with indices that the JIT compiler cannot resolve may remain as memory
loads and stores in compiled code. Representing the stack as a fixed-size
virtualizable array allows the JIT compiler to optimize individual elements, but
makes the virtualizable state depend on the array's allocated capacity rather
than the active stack depth. Neither approach directly supports optimizing a
bounded portion of a shared stack with dynamically varying depth while
preserving the remaining elements in memory.

We study this problem using Forth because it does not guarantee static stack-use
information. Stack-effect annotations are not enforced, \code{EXECUTE} permits
indirect calls whose stack effects may not be statically known, and extensions
such as \code{PICK} and \code{ROLL} allow data-dependent access to deeper stack
elements. Moreover, most Forth primitives manipulate the shared data stack,
making its representation critical to the JIT compiler's ability to eliminate
memory operations. These characteristics make Forth a demanding test case for
stack optimizations that do not rely on static stack-use information. Forth also
provides established benchmarks and native compilers for performance comparison.

We propose the \emph{stack fragment layout}, which exposes a bounded portion of
a shared data stack to the meta-tracing JIT compiler without requiring static
stack-use information. The layout divides the stack into a fixed-width \emph{top
  fragment} and a shared \emph{spill array}. The top fragment stores the
uppermost stack cells in fixed interpreter fields, allowing the JIT compiler to
track their values within a trace, while the remaining cells reside in the spill
array. A decoding function defines how these two components represent the
logical stack. Since the fragment has a fixed width, the number of stack cells
exposed to the compiler remains bounded regardless of the stack depth. At trace
exits requiring state reconstruction, only the top fragment and a small amount
of associated control state, such as its occupancy, need to be reconstructed,
provided that the remaining stack cells are already stored in the spill array.

The layout preserves the semantics of the modeled stack operations across word
calls, provided that the spill array has sufficient capacity. In particular, a
callee can access values left by its caller without requiring a separate
data-stack frame or statically known stack effects. We also introduce a
\emph{call-entry policy} that controls the number of cells held in the
top fragment at word entry. We consider two policies: \emph{call-entry
  normalization}, which uses a fixed target occupancy, and \emph{adaptive
  entry}, which selects a target occupancy for each word. We implement the
layout in \emph{\system{}}, an RPython-based Forth interpreter, and in
\emph{RPyFactor}, an interpreter for a subset of Factor. The latter demonstrates
the applicability of the layout to a language with declared stack effects,
without relying on those declarations.

We evaluate \system{} using eighteen Shootout kernels, fourteen Are We Fast Yet
benchmarks, and six Appbench applications on two x86-64 machines, and RPyFactor
using twenty ported kernels. To isolate the benefit of exposing stack values to
the JIT, our primary comparison enables or disables top-fragment virtualization
while keeping the physical stack layout and call-entry policy unchanged. The
non-virtualized configuration serves as a conventional stack cache, whose
accesses may remain as memory operations in compiled traces. We further
investigate how performance is affected by fragment width, stack-access
patterns, call-entry policies, and the cost of moving values between the
fragment and spill. We also examine trace size and the effectiveness of a
separate floating-point top fragment.

This paper makes three contributions:

\begin{enumerate}
\item \emph{Stack fragment layout.}
  We introduce a stack representation that exposes a fixed-width top fragment
  of a call-shared data stack to a meta-tracing JIT compiler while retaining
  the remaining cells in a shared spill array.
  The layout requires no static stack-effect information and bounds the
  stack state reconstructed at trace exits independently of the total stack
  depth, under stated assumptions.
  We provide refinement arguments for the modeled stack operations and
  characterize the requirements for adopting the layout in other interpreters
  (\cref{sec:layouts,thm:refinement,lem:exit-bound,sec:adoption,sec:other-vms}).

\item \emph{Interpreter implementations.}
  We implement the layout in \system{}, an RPython-based Forth interpreter
  covering the Forth Core word set, including integration with word calls,
  interpreter dispatch, and a floating-point stack.
  We also implement the layout in RPyFactor, an interpreter for a subset of
  Factor, to demonstrate its applicability beyond Forth without relying on
  declared stack effects
  (\cref{sec:implementation,sec:factor}).

\item \emph{Experimental evaluation.}
  We isolate the performance benefit of exposing the top fragment to the JIT
  while keeping the stack layout and call-entry policy fixed.
  Across the evaluated suites, JIT exposure yields geometric-mean speedups
  of \(1.44\)--\(2.11\times\) in \system{} and \(1.42\)--\(1.56\times\)
  in RPyFactor.
  We further examine fragment configurations, call-entry policies, and
  floating-point stack optimization, and compare \system{} with established
  native Forth compilers
  (\cref{sec:rq1-annotations,sec:float-fragment,sec:factor,sec:engines}).
\end{enumerate}

\Cref{sec:background} gives background, \cref{sec:layouts} presents the
layout, \cref{sec:implementation} its implementation, and
\cref{sec:evaluation} the evaluation.  \Cref{sec:discussion} discusses the
results and threats to validity, \cref{sec:related} relates the work to prior
designs, and \cref{sec:conclusion} concludes.  The appendices give the state
model with proof sketches and supplementary measurements.

\section{Background}
\label{sec:background}

\subsection{Forth}
\label{sec:forth-background}

A Forth system is organized around a \emph{dictionary} of words and two stacks
(three with the optional floating-point words).  Its \emph{outer} interpreter reads whitespace-separated
tokens from the input and looks each one up in the dictionary.  It executes
the word it finds or, if the lookup fails and the token is a number, pushes it
on the data stack.  The colon compiler \code{:} puts the system into
compilation state, where words and numbers are compiled into a new definition
rather than executed, except for \emph{immediate} words, until \code{;} closes it.  In a threaded
system, the \emph{inner} interpreter runs such a definition by dispatching its
sequence of word references~\cite{ertl03,vmgen02,koopman89}.

Words take operands from the data stack and leave results there.  The
return stack holds control state such as return addresses and loop parameters,
and can also hold temporary values.  \code{PICK} copies and \code{ROLL} moves
a cell at a depth supplied on the data stack, and \code{EXECUTE} invokes the
word identified by an execution token on that stack~\cite{forth2012}.  A
stack-effect comment describes what a word consumes and produces, but no
implementation is required to check or know it before execution.

These properties shape how Forth is implemented.  Interpretive systems attack
dispatch cost with threading techniques and superinstructions, and they attack
memory traffic by keeping the topmost cells in registers, a technique known as
stack caching~\cite{ertl95,ertl04,ertl05}.  Native compilers such
as VFX Forth and SwiftForth go further and compile each definition to machine
code, optimizing stack traffic over a region of the
program~\cite{ertl96,vfxmanual,swiftmanual}.  Both traditions keep the shared
data stack; they differ in how much of it they can hold outside memory at a
given moment.

\subsection{One data stack crosses word calls}
\label{sec:shared-stack}

We write a logical data stack as a finite sequence
\(\stackseq = [a_0,a_1,\ldots,a_{|\stackseq|-1}]\) ordered from bottom to
top.  Push appends a value; a successful pop removes
\(a_{|\stackseq|-1}\); and, for \(0\le k<|\stackseq|\), a depth-\(k\) read
returns
\(a_{|\stackseq|-1-k}\).  A control-only call or return preserves \(\stackseq\); the
callee operates on the same sequence as its caller.  In
\cref{fig:callshared}, \code{SUM3} replaces the top three cells with their sum.
Exception recovery differs from ordinary return.  \code{CATCH} saves the stack
depths, and a \code{THROW} it handles restores them before execution resumes
after that \code{CATCH}~\cite{forth2012}.  Implementations impose finite stack capacities; shared
stack semantics does not require an infinite backing store.

\begingroup
\tikzset{
  state/.style={draw=figInk!90,minimum width=10mm,minimum height=5.5mm,font=\small},
  shared/.style={state,fill=figFill,text=black},
  arg/.style={state,draw=figEmph,fill=figEmph!25,line width=1pt},
  phase/.style={font=\scriptsize\bfseries,align=center,text=black},
  note/.style={font=\scriptsize,align=center,text=black},
  callee/.style={draw=figInk!75,dashed,rounded corners=3pt,fill=figFill!35},
  transition/.style={-{Stealth[length=1.8mm]},draw=figInk,line width=.75pt}
}
\begin{figure}[!t]
\centering
\begin{tikzpicture}[x=.9cm,y=1cm,transform shape,scale=0.9]
\draw[callee] (2.55,-1.56) rectangle (8.25,2.76);
\node[phase] at (0,2.55) {caller before call};
\matrix (before) [matrix of nodes,row sep=-\pgflinewidth,anchor=south] at (0,-1.5) {
  |[arg]| \(30\) \\
  |[arg]| \(20\) \\
  |[arg]| \(10\) \\
  |[shared]| \(q\) \\
  |[shared]| \(p\) \\
};

\node[phase] at (3.6,2.55) {\code{SUM3} entry};
\matrix (entry) [matrix of nodes,row sep=-\pgflinewidth,anchor=south] at (3.6,-1.5) {
  |[arg]| \(30\) \\
  |[arg]| \(20\) \\
  |[arg]| \(10\) \\
  |[shared]| \(q\) \\
  |[shared]| \(p\) \\
};

\node[phase] at (7.2,2.55) {\code{SUM3} exit};
\matrix (exit) [matrix of nodes,row sep=-\pgflinewidth,anchor=south] at (7.2,-1.5) {
  |[arg]| \(60\) \\
  |[shared]| \(q\) \\
  |[shared]| \(p\) \\
};

\node[phase] at (10.8,2.55) {caller after return};
\matrix (after) [matrix of nodes,row sep=-\pgflinewidth,anchor=south] at (10.8,-1.5) {
  |[arg]| \(60\) \\
  |[shared]| \(q\) \\
  |[shared]| \(p\) \\
};

\draw[transition] (0.88,0) -- node[above,note] {call} (2.72,0);
\draw[transition] (4.48,0) -- node[above,note] {pop three\\push \(60\)} (6.32,0);
\draw[transition] (8.08,0) -- node[above,note] {return} (9.92,0);
\foreach \m in {before,entry,exit,after}{\node[note,above=1mm of \m-1-1] {top};}
\end{tikzpicture}
\caption{A word call and return leave the call-shared data stack in place.  \code{SUM3} consumes the three blue cells that the caller pushed and leaves \(60\) in their place.  The grey cells \(p,q\) lie below them and stay unchanged.  The four columns are successive states of one logical stack, not per-call frames, and the dashed box marks those inside \code{SUM3}.}
\label{fig:callshared}
\end{figure}
\endgroup

\subsection{Meta-tracing and interpreter state}
\label{sec:metatracing}

A tracing JIT compiler observes a running program, detects a hot loop, and
records the operations one iteration actually
performs~\cite{bala00,gal06,gal09}. The recorded trace is
linear: each conditional branch whose outcome is not already known becomes a
\emph{guard}, an operation that checks the recorded outcome in the compiled
code and leaves the trace when it differs.  A \emph{meta-tracing} compiler
applies this to an interpreter rather than to the program it
runs~\cite{bolz09,bolztratt15}. The virtual machine (VM) implementer marks the
interpreter's dispatch loop and the variables that identify a position in
the user program, typically the instruction pointer and the current word.
Their values are known at each recorded dispatch, so dispatch, decoding, and
the lookups that depend on them fold away.  A guard that fails often gets a
bridge that compiles the alternate path.  The optimizer additionally specializes the trace on
the loop's invariants~\cite{ardo12}.

Leaving a trace is what constrains the state representation.  When a guard
fails and has no bridge, execution resumes in the interpreter, so the compiler
must be able to rebuild the interpreter state needed to resume, the same obligation a deoptimizing compiler has towards its
interpreter~\cite{holzle92}.  It therefore attaches to each guard a
\emph{guard-exit record}.  RPython's resume data describe how to rebuild the
state, and the guard's \emph{failargs} are the live values the rebuild reads.
The more interpreter state a trace carries as compiler values, the more failargs
its guards can carry, which is why \cref{sec:rq2-traces} counts them.

RPython is the language and toolchain this compiler comes
with~\cite{bolz09,bolztratt15}.  An interpreter is written in a restricted
subset of Python and translated ahead of time to C.  The toolchain integrates
a garbage collector and generates the tracing JIT compiler.  The implementer
writes no optimizer but guides tracing and specialization with hints.

Two mechanisms concern state.  A \emph{virtual} is an object whose allocation the optimizer can
defer.  The trace keeps its fields as static single assignment (SSA) values and
materializes the object when compiled execution or interpreter resumption needs
it~\cite{bolz11}.  A \emph{virtualizable} is a long-lived object, such as an
interpreter frame, with fields that the implementer marks for tracking as
trace values.  The runtime synchronizes those fields when execution outside the
trace needs them~\cite{rpythonvirtualizables}.

What a virtualizable exposes is decided one field at a time.  A fixed-size
array field declares all its elements at once, as in \code{frame[*]}.  An
element becomes an individual trace value only when the trace knows the index
of the access.  For a dynamic index, RPython guards that it equals the value
seen while tracing~\cite{rpythonvirtualizables}.\footnote{\url{https://github.com/pypy/pypy/blob/51a08c6f0c21ef6a9d7a794db6d3f76fb7b0b558/rpython/jit/metainterp/pyjitpl.py\#L1221-L1225}}  A bytecode frame of statically
known size fits this scheme, because each activation has a fixed set of local
and operand slots.

A Forth data stack has finite capacity but no such per-call bound, as the next
section shows.

\section{Problem and Stack Fragment Layout}
\label{sec:layouts}
\label{sec:problem}

\Cref{sec:background} suggests three ways to hold a call-shared stack in
RPython, and none fits.  Keeping it in one array lets the stack pointer select
each cell, so exposing a cell needs an index guard, including for a runtime
depth such as \code{PICK}'s.  Declaring the whole array as a virtualizable
field ties the exposed state to the capacity a build reserves.  Giving each
call a private frame, as bytecode VMs do, breaks access to the caller's cells
(\cref{sec:shared-stack}).  We need a representation that exposes a bounded
top part and keeps the rest recoverable.  We propose the stack fragment layout,
which exposes a fixed-width top fragment of
individually identifiable cells above a shared spill array.

\subsection{Stack fragment layout}

The stack fragment layout exposes a bounded top fragment and leaves deeper cells
in ordinary shared memory.  A physical state has three ordered regions,
\[
 q=\langle S,F,T\rangle,\qquad
 \decode(q)=S\mathbin{\|}F\mathbin{\|}T,\qquad
 |F|\le f,\quad |T|=\min(2,|F|+|T|),\quad W=f+2 .
\]
Here \(S\) is the occupied prefix of the shared spill, \(F\) is the fixed-frame
region, and \(T\) contains up to two scalar top cells, all ordered from bottom
to top.  The scalar region holds the topmost \(\min(2,|F|+|T|)\) active cells.
The frame width \(f\) is fixed for a build, so the exposed top fragment has \(W\)
locations while the logical stack can grow beyond it.  Pushes and pops move
cells across region boundaries, and an indexed access may reach any of the
three.

A word call creates no private data frame.  Calls and returns preserve the logical stack.  Optional call-entry
normalization spills all but the top \(c\) exposed cells.  Tail calls and
ordinary returns preserve the split.
For the current exposed sequence \(U=F\mathbin{\|}T\), write
\(U=P\mathbin{\|}K\) with \(|K|=\min(c,|U|)\).  Normalization, for
\(0\le c\le W\), is then
\[
 N_c(S,U)=(S\mathbin{\|}P,K).
\]
The move needs room for \(P\) in the spill, which the capacity-reserved
domain of \cref{sec:layout-correctness} guarantees.  The decoded stack is
unchanged and at most \(c\) cells stay active, with
\(K\) split between frame and scalar positions.  Normalization never refills
the top fragment from the spill when fewer than \(c\) cells are active already.
\emph{Materialization} is \(N_0\), which empties the top fragment into the spill.
The evaluated default uses \(f=8\) and \(c=2\).

\subsection{Adaptive entry}
\label{sec:adaptive}

Adaptive entry keeps the same top fragment and shared spill.  Where a call
normalizes, it uses a per-word target \(c_w\) in place of \(c\):
\[
 \operatorname{enter}_w(S,U)=N_{c_w}(S,U),\qquad 2\le c_w\le W.
\]
A linear scan estimates the number of entry cells required, stopping at
\code{EXIT} or an unknown effect.  It adds two for bodies of at least 48
instructions.  Among classes in \(\{2,4,8,10\}\) that fit within \(W\), it
selects the smallest that meets this estimate, or the largest available.
\Cref{sec:rq3} tests this policy.

\subsection{Operations across the layout}
\label{sec:layout-ops}

\begingroup
\tikzset{
  cell/.style={draw=figInk,minimum width=10mm,minimum height=5mm,font=\scriptsize,inner sep=1pt},
  scalar/.style={cell,draw=figEmph,fill=figEmph!40,line width=.8pt},
  frame/.style={cell,draw=figEmph,fill=figEmph!16},
  spill/.style={cell,draw=figInk!75,fill=figFill,minimum width=8mm},
  inactive/.style={cell,draw=figEmph!85,dashed},
  phase/.style={font=\scriptsize\bfseries,align=center,text=black},
  note/.style={font=\scriptsize,align=center,text=black},
  region/.style={font=\scriptsize,text=black,anchor=east},
  transition/.style={-{Stealth[length=1.7mm]},draw=figInk,line width=.75pt},
  motion/.style={-{Stealth[length=1.7mm]},draw=figEmph,line width=.8pt,dashed},
  spillmotion/.style={-{Stealth[length=1.7mm]},draw=figAcc,line width=.8pt,dashed}
}
\begin{figure}[!t]
\centering
\begin{tikzpicture}[x=.75cm,y=1cm]
\node[phase] at (3.15,2.15) {full-fragment push};
\node[phase] at (11.7,2.15) {pop};

\node[phase] at (1.3,1.8) {before};
\node[phase] at (5.0,1.8) {after};
\node[phase] at (9.4,1.8) {before};
\node[phase] at (14.0,1.8) {after};

\node[scalar] (pb0) at (1.3,1.35) {\(A\)};
\node[scalar] (pb1) at (1.3,.85) {\(B\)};
\node[frame]  (pb2) at (1.3,.25) {\(C\)};
\node[frame]  (pb3) at (1.3,-.25) {\(D\)};
\node[region] at (.1,1.1) {scalar};
\node[region] at (.1,0) {fixed frame};
\node[spill] (pbs0) at (.8,-1.05) {\(F\)};
\node[spill] (pbs1) at (1.8,-1.05) {\(E\)};
\node[region] at (.1,-1.05) {spill};

\node[scalar] (pa0) at (5.0,1.35) {\(x\)};
\node[scalar] (pa1) at (5.0,.85) {\(A\)};
\node[frame]  (pa2) at (5.0,.25) {\(B\)};
\node[frame]  (pa3) at (5.0,-.25) {\(C\)};
\node[spill] (pas0) at (4.0,-1.05) {\(F\)};
\node[spill] (pas1) at (5.0,-1.05) {\(E\)};
\node[spill,draw=figAcc,line width=.8pt] (pas2) at (6.0,-1.05) {\(D\)};

\draw[transition] (pb0.east) ++(.3,-.6) -- node[above,note] {push \(x\)} ++(1.75,0);
\draw[spillmotion] (pb3.east) -- (2.75,-.25) -- (2.75,-1.75)
  -- node[below,note,pos=.5] {deepest cell spills} (6.0,-1.75) -- (pas2.south);

\node[scalar] (qb0) at (9.9,1.35) {\(x\)};
\node[scalar] (qb1) at (9.9,.85) {\(A\)};
\node[frame]  (qb2) at (9.9,.25) {\(B\)};
\node[frame]  (qb3) at (9.9,-.25) {\(C\)};
\node[region] at (8.2,1.1) {scalar};
\node[region] at (8.2,0) {fixed frame};
\node[spill] at (8.9,-1.05) {\(F\)};
\node[spill] at (9.9,-1.05) {\(E\)};
\node[spill] at (10.9,-1.05) {\(D\)};
\node[region] at (8.2,-1.05) {spill};

\node[scalar] (qa0) at (14.0,1.35) {\(A\)};
\node[scalar] (qa1) at (14.0,.85) {\(B\)};
\node[frame]  (qa2) at (14.0,.25) {\(C\)};
\node[inactive] (qa3) at (14.0,-.25) {};
\node[spill] at (13.0,-1.05) {\(F\)};
\node[spill] at (14.0,-1.05) {\(E\)};
\node[spill] at (15.0,-1.05) {\(D\)};

\draw[transition] ([xshift=2mm]qb0.east) -- node[above,note] {pop, returns \(x\)} ([xshift=-2mm]qa0.west);
\draw[motion] (qb2.east) -- node[below,note] {frame top\\moves up} (12.9,.25) |- (qa1.west);
\end{tikzpicture}
\caption{Full-fragment push and pop for \(W=4\), with two scalar and two fixed-frame positions; the evaluated default has eight frame positions.  Scalar and frame cells are shown topmost first, and the spill from bottom (left) to top (right).  Dashed arrows show cells that change region, in red when a cell moves to the spill.  After the pop, \(C\) is the only occupied frame cell, and the dashed slot is inactive, not zeroed.}
\label{fig:layout}
\end{figure}
\endgroup

A push puts the new value in the top scalar field, and each cell below it
becomes one position deeper.  When the top fragment is already full, it also moves the
deepest exposed cell out to the spill to make that room, which is called
\emph{full-fragment push} here.  A pop removes the top
active cell, or reads and removes the last spill cell when the top fragment is empty,
and it never refills the top fragment from the spill.  An indexed read or write selects a scalar
field, a fixed-frame position, or a spill offset. \Cref{fig:layout}
illustrates a full-fragment push and a pop with two scalar and two frame
positions, whereas the evaluated default has eight frame positions.

\subsection{Correctness}
\label{sec:layout-correctness}

Two properties need an argument.  The split representation has to behave like
the single stack of \cref{sec:shared-stack}, and the stack representation's
share of an exit's state has to follow the top fragment's width rather than the
stack's depth.  Both use the projected state \(q=\langle S,U\rangle\), where
\(U=F\mathbin{\|}T\), \(\decode(q)=S\mathbin{\|}U\), and \(|U|\le W\).
Write \(\wmax=W\).  The model assumes \(|S|+|U|\le\dmax\le\bspill\), where
\(\dmax\) bounds logical depth and \(\bspill\) is spill capacity.  This reserves
room to spill the entire stack and excludes near-capacity implementation
states.

\begin{restatable}[Stack-sequence refinement]{theorem}{thmRefinement}
\label{thm:refinement}
Let \(q=\langle S,U\rangle\) satisfy \(|U|\le\wmax\) and
\(|S|+|U|\le\dmax\le\bspill\), and use the operation rules of
\cref{app:operations}.  For every successful push, pop, indexed write, or reset
taking \(q\) to \(q'\), \(\decode(q')\) equals the result of the corresponding
abstract stack operation on \(\decode(q)\); pop also returns the same value.
Indexed reads at valid depths and depth queries return the corresponding
abstract results.  Any sequence-preserving policy step leaves \(\decode(q)\)
unchanged.  Such steps include normalization, materialization, and control-only calls and returns.
\end{restatable}

The three-region simulation in \cref{app:three-regions} transfers refinement
to the concrete layout for successful modeled operations within the
capacity-reserved domain.

\begin{restatable}[Conditional reconstruction bound]{lemma}{lemExitBound}
\label{lem:exit-bound}
Assume that the top fragment has at most \(\wmax\) individually identifiable
locations, that all current deeper-cell values already reside in a shared
spill described by \(O(1)\) references and indices, that metadata per
exposed location has constant size, and that these values and the occupied
top-fragment locations determine the representation at the exit.  Then reconstructing the data-stack representation at an optimized-code exit
requires at most \(\wmax\) current top-fragment cell values plus \(O(1)\) control
values, such as the occupancy and the spill index, with \(O(\wmax)\) representation-derived metadata independent of
\(\bspill\).
\end{restatable}

The lemma bounds only the stack representation's share of an exit, under its
spill assumption.  Whole guard-exit records also hold other live state, and
\cref{sec:rq2-traces} counts all their values at fixed \(W\), so the
stack-specific bound remains analytical.  \Cref{app:layout} gives proof
sketches.  Refinement composes over the disjoint floating-point top fragment, and the
bound adds both widths when each meets the assumptions of \cref{lem:exit-bound}
(\cref{app:two-stacks}).

\FloatBarrier
\section{Implementation}
\label{sec:realization}
\label{sec:implementation}
\label{sec:rpython}
\label{sec:integration}

\begingroup
\tikzset{
  acell/.style={draw=figInk!90,fill=figFill,minimum width=7mm,minimum height=7mm,font=\small},
  memcell/.style={acell,draw=figAcc,line width=1.1pt},
  wscalar/.style={draw=figEmph,fill=figEmph!40,line width=.9pt,minimum width=9mm,minimum height=7mm,font=\small},
  wframe/.style={draw=figEmph,fill=figEmph!16,minimum width=5mm,minimum height=7mm},
  wspill/.style={draw=figAcc,dashed,fill=figAcc!14,minimum width=13mm,minimum height=7mm,font=\footnotesize},
  depth/.style={draw=figEmph,rounded corners=1.5pt,fill=white,font=\footnotesize,inner sep=1.5pt},
  tl/.style={font=\footnotesize\ttfamily,anchor=north west,text width=60mm,inner xsep=1.5mm,
             inner ysep=0pt,minimum height=4.6mm,text depth=.25ex,align=left},
  tmem/.style={tl,fill=figAcc!25},
  tgrd/.style={tl,fill=figEmph!25},
  tval/.style={tl},
  tnote/.style={tl,font=\footnotesize\itshape,text=black},
  tbox/.style={draw=figInk!75,fill=figCode,rounded corners=2pt,inner sep=1.2mm},
  head/.style={font=\small\bfseries,text=black},
  tag/.style={font=\footnotesize,text=black,align=center},
  tracehead/.style={font=\footnotesize\bfseries,text=black,anchor=south east},
  memarrow/.style={-{Stealth[length=1.8mm]},draw=figAcc,line width=.95pt,dashed},
  valarrow/.style={-{Stealth[length=1.8mm]},draw=figEmph,line width=1pt},
  grdarrow/.style={-{Stealth[length=1.8mm]},draw=figEmph,line width=.8pt},
  arrow/.style={-{Stealth[length=1.8mm]},draw=figInk,line width=.8pt}
}
\begin{figure}[!t]
\centering
\begin{tikzpicture}[x=1cm,y=1cm,scale=0.9,transform shape]
\node[head] at (3.0,2.75) {one array};
\foreach \i/\v in {0/,1/,2/{\(a\)},4/,5/}{\node[acell] at (0.875+\i*0.75,1.5) {\v};}
\node[memcell] (b) at (3.125,1.5) {\(b\)};
\draw[arrow] (3.875,0.86) node[tag,below right,xshift=-1.5mm] {\code{sp} (next free)} -- (3.875,1.12);
\node[tmem] (l1) at (0.1,-0.35) {i1 = getarrayitem(p, i\_sp - 1)};
\node[tval] (l2) at (l1.south west) {i2 = int\_mul(i1, i1)};
\node[tmem] (l3) at (l2.south west) {setarrayitem(p, i\_sp - 1, i2)};
\node[tracehead,anchor=south west] at ([yshift=1.2mm]l1.north west) {trace of \code{DUP *}};
\begin{scope}[on background layer]
\node[tbox,fit=(l1)(l3)] {};
\end{scope}
\draw[memarrow] (b.south) -- node[tag,left] {load} (b.south |- l1.north);
\draw[memarrow] (l3.east) -- ++(.35,0) |- node[tag,pos=.75,above] {store} (3.3,2.15) -- (3.3,1.5 |- b.north);
\node[head] at (10.5,2.75) {top fragment \(+\) spill};
\node[wspill] at (8.2,1.5) {\code{spill}};
\foreach \i in {0,...,3}{\node[wframe] at (9.3+\i*0.55,1.5) {};}
\draw[decorate,decoration={brace,amplitude=3pt},draw=figEmph] (9.05,1.92) -- (11.2,1.92) node[tag,midway,above=2pt] {\code{frame}};
\node[wscalar] (t1) at (11.9,1.5) {\code{t1}};
\node[wscalar] (t0) at (12.9,1.5) {\code{t0}};
\node[tag,above=1mm of t0] {top};
\node[depth] (cd) at (8.2,0.45) {\code{cache\_depth} \(=2\)};
\node[tgrd,text width=48mm] (g1) at (7.6,-0.35) {guard(cache\_depth == 2)};
\node[tval,text width=48mm] (g2) at (g1.south west) {i2 = int\_mul(i1, i1)};
\node[tnote,text width=48mm] (g3) at (g2.south west) {no load or store: i1 is \code{t0}};
\node[tracehead] at ([yshift=1.2mm]g1.north east) {trace of \code{DUP *}};
\begin{scope}[on background layer]
\node[tbox,fit=(g1)(g3)] {};
\end{scope}
\draw[grdarrow] (cd.south) -- (cd.south |- g1.north);
\draw[valarrow] (t0.south) -- ++(0,-0.3) -| node[tag,pos=.75,right,text=figEmph] {value \code{i1}} ([xshift=7mm]g2.east) -- ([xshift=-16mm]g2.east);
\end{tikzpicture}
\caption{Schematic trace of \code{: SQUARE DUP * ;} inside a hot loop.  Both
representations are drawn from bottom (left) to top (right).  With one array,
the trace loads and stores the top cell \(b\) at \code{sp} minus one (dashed,
red).  With the top fragment, the top
cells are the fields \code{t0}, \code{t1}, and \code{frame}.  One guard on
\code{cache\_depth} fixes their positions, and the trace uses
\code{t0} directly as the value \code{i1} (blue), so the multiplication needs
no load or store.}
\label{fig:vm-overview}
\end{figure}
\endgroup

\system{} implements the fragment layout in an RPython interpreter for Forth
with two levels.  The outer interpreter compiles source into a dictionary of
words, each a thread of word references with a parallel array of literals.
The inner interpreter executes these threads: it runs primitives and enters
colon words by call or tail transfer.  The RPython toolchain translates the
interpreter into a native executable whose generated meta-tracing JIT compiler
compiles the hot dispatch paths (\cref{fig:vm-overview}).\footnote{The outer
  interpreter also rewrites threads as it builds them.  It splices eligible
  colon bodies of up to 32 instructions and relocates their branches,
  specializes selected defining-word patterns, bounds recursive expansion, and
  turns final calls into tail transfers.  The RQ1 annotation controls share
  these rewrites and differ only in the fields they expose.  The engine
  comparison includes the rewrites' effect.}

\subsection{Architecture overview}
\label{sec:architecture}

\begingroup
\tikzset{
  abox/.style={draw=figInk,rounded corners=2pt,font=\footnotesize,align=center,
               fill=white,inner sep=1.5mm},
  region/.style={draw=figInk!75,fill=figFill!35,rounded corners=3pt,inner sep=1.6mm},
  objregion/.style={draw=figInk!75,fill=figFill!35,rounded corners=3pt,inner sep=1.6mm},
  rtitle/.style={font=\footnotesize\bfseries,text=black,anchor=north west},
  sc/.style={draw=figEmph,fill=figEmph!40,line width=.8pt,minimum width=8mm,minimum height=6mm,font=\footnotesize},
  fr/.style={draw=figEmph,fill=figEmph!16,minimum width=4.5mm,minimum height=6mm},
  sp/.style={draw=figAcc,dashed,fill=figAcc!14,minimum width=30mm,minimum height=6mm,font=\footnotesize},
  ctl/.style={draw=figInk,fill=figFill,minimum width=14mm,minimum height=6mm,font=\footnotesize,align=center},
  ctlrow/.style={draw=figInk,fill=figFill,minimum width=28mm,minimum height=4.5mm,font=\footnotesize,inner sep=1pt},
  tag/.style={font=\footnotesize,text=black,align=center},
  use/.style={-{Stealth[length=1.6mm]},draw=figInk,line width=.7pt,shorten >=1.2pt},
  both/.style={{Stealth[length=1.6mm]}-{Stealth[length=1.6mm]},draw=figInk,line width=.7pt,shorten >=1.2pt,shorten <=1.2pt},
  ref/.style={{Circle[length=1.3mm]}-{Stealth[length=1.6mm]},draw=figInk,line width=.7pt,shorten >=1.2pt},
  ptr/.style={draw=figInk,fill=white,minimum height=4.5mm,font=\footnotesize,inner sep=2pt},
  tc/.style={draw=figInk!90,fill=white,minimum width=9.5mm,minimum height=6mm,font=\footnotesize,inner sep=1pt},
  code/.style={draw=figInk!75,rounded corners=2pt,fill=figCode,font=\footnotesize\ttfamily,align=left,inner sep=1.5mm},
  loopback/.style={-{Stealth[length=1.6mm]},draw=figInk,line width=.7pt},
  move/.style={{Stealth[length=1.4mm]}-{Stealth[length=1.4mm]},draw=figAcc,line width=.8pt,dashed,shorten >=1.2pt,shorten <=1.2pt}
}
\begin{figure}[tb]
\centering
\begin{tikzpicture}[x=1cm,y=1cm,scale=0.86,transform shape]
\node[rtitle,anchor=east] (vtitle) at (8.55,7.62) {inner interpreter (virtualizable)};
\node[abox,anchor=west] (outer) at (-1.45,8.75) {\textbf{outer interpreter}\\parse, compile, rewrite};
\node[code,anchor=west] (src) at (3.6,8.75) {: W DUP * S>F F+ ;};
\node[tag,anchor=west] at (src.east) {Forth program};
\draw[use] (src.west) -- (outer.east);
\foreach \i/\w in {0/DUP,1/*,2/S>F,3/F+,4/EXIT}{\node[tc] (tc\i) at (-0.95+\i*0.95,6.6) {\code{\w}};}
\node[tag,anchor=north west] at ([yshift=-1mm]tc2.south west) {code thread};
\draw[use] (tc0.north |- outer.south) -- node[tag,right,pos=.35] {threads, literals} (tc0.north);
\draw[use] (tc1.south |- 0,5.6) node[tag,below] {\code{ip}} -- (tc1.south);
\node[code] (disp) at (5.45,5.15) {%
  loop:\\
  \ \ w := thread[ip]; ip += 1\\
  \ \ if w is a primitive:\\
  \ \ \ \ pop; compute; push\\
  \ \ else: enter w};
\draw[use] (tc1.north) |- (disp.north |- 0,7.2) -- node[tag,right,pos=.4] {fetch \code{thread[ip]}} (disp.north);
\draw[loopback] ([yshift=2mm]disp.south west) to[out=180,in=180,looseness=1.3] ([yshift=-2.5mm]disp.north west);
\node[ptr] (fsf) at (8.25,3.0) {\code{fstack}};
\node[font=\footnotesize,text=black,align=left,anchor=center] (ctl) at (-0.15,2.85)
  {\textbf{other fields}\\[1pt]
   \textbullet\ return-stack pointer\\
   \textbullet\ call-stack pointer\\
   \textbullet\ loop index\\
   \textbullet\ locals pointer, \dots};
\foreach \i in {0,...,3}{\node[fr] (if\i) at (2.75+\i*0.5,3.0) {};}
\node[sc] (it1) at (4.9,3.0) {\code{t1}};
\node[sc] (it0) at (5.75,3.0) {\code{t0}};
\node[tag,anchor=north] (itag) at (4.2,2.62) {integer top fragment: \code{frame}, \code{cache\_depth}};
\begin{scope}[on background layer]\node[region,fit=(vtitle)(tc0)(disp)(fsf)(ctl)(if0)(it0)(itag)] (vf) {};\end{scope}
\node[rtitle,anchor=west,align=left] (ftitle) at (9.65,4.0) {floating-point top fragment\\(heap object)};
\foreach \i in {0,...,3}{\node[fr] (ff\i) at (10.0+\i*0.5,3.0) {};}
\node[sc] (ft1) at (12.2,3.0) {\code{ft1}};
\node[sc] (ft0) at (13.05,3.0) {\code{ft0}};
\node[tag,anchor=north] (ftag) at (11.5,2.62) {\code{fframe}, \code{fdep} (promoted)};
\begin{scope}[on background layer]\node[objregion,fit=(ftitle)(ff0)(ft0)(ftag)] (fo) {};\end{scope}
\node[sp] (isp) at (4.2,0.7) {integer spill};
\node[sp] (fsp) at (11.5,0.7) {floating-point spill};
\node[tag] at (7.85,0.7) {spill arrays (heap)};
\draw[both] (disp.south -| it0.north) -- node[tag,left] {integer\\pop/push} (it0.north);
\draw[both] (disp.east) -| node[tag,left,pos=.88] {float\\pop/push} (fsf.north);
\draw[ref] (fsf.east) -- (fsf.east -| fo.west);
\draw[move] (isp.north) -- node[tag,right] {full push, empty pop} (isp.north |- vf.south);
\draw[move] (fsp.north) -- node[tag,left] {full push, empty pop} (fsp.north |- fo.south);
\end{tikzpicture}
\caption{The two top fragments of \system{}.  The outer interpreter compiles
the Forth program into threads of word references with literal arrays, and the
dispatch loop of the inner interpreter runs a thread one word at a time.  A stack primitive uses one top fragment or both, and the top fragments only hold values.  The inner interpreter is the virtualizable
object and holds the integer top fragment with the other interpreter state.  The
floating-point top fragment of the same shape lives in a separate, ordinary heap
object, reached through the immutable field \code{fstack}.  A word such as
\code{S>F} pops from one top fragment and pushes onto the other.  Each top fragment has its
own spill, which a pop reads only when the top fragment is empty.}
\label{fig:architecture}
\end{figure}
\endgroup

\Cref{fig:architecture} shows where the two top fragments live and which operations
use them.  The outer interpreter compiles each definition of the Forth program
into a thread, and the dispatch loop of the inner interpreter runs threads one
word at a time.  The inner interpreter is one long-lived object, and RPython treats
it as the virtualizable frame of every trace.  It holds the control state,
such as the return-stack and loop pointers, and the integer top fragment.  The
floating-point top fragment has the same shape but lives in a separate, ordinary heap
object, which the interpreter reaches through an immutable field.  The
optimizer treats its fields like those of any heap object rather than as
virtualizable state.  Each top fragment has its own
spill array in the heap, and neither spill is virtualizable.

Each primitive names the stack it uses.  Integer words such as \code{DUP} and
\code{+} use only the integer top fragment, and floating-point words such as
\code{F+} and \code{FDUP} use only the floating-point top fragment.  Words that cross
the stacks pop from one top fragment and push onto the other.  \code{S>F} converts an
integer and pushes it onto the floating-point stack, \code{F<} leaves its flag on the
integer stack, and \code{F@} takes its address from the integer stack.  The
top fragments are also treated differently at calls and exceptions.  A colon call
normalizes only the integer top fragment (\cref{sec:layouts}).  For a colon word,
\code{CATCH} snapshots the integer top fragment but saves only the floating-point
depth, which \code{THROW} restores.  A trace that does not touch the
floating-point stack therefore carries none of its fields, whereas a trace that
does is specialized on the promoted floating-point occupancy.

\subsection{The implemented stack layout}
\label{sec:window-implementation}

\system{} instantiates the fragment layout of \cref{sec:layouts} with a frame
width of \(f=8\) and an entry target of \(c=2\).  Each region becomes state in
the long-lived inner interpreter rather than an object allocated at a word
call.  The scalar region is the fields \code{t0} and \code{t1}, the fixed frame
is \code{frame[*]} with its eight positions, and the spill is a shared array
addressed through \code{spill\_ptr}, while \code{cache\_depth} holds how many
cells the top fragment currently has.  The exposed build declares everything but the spill array
virtualizable, together with the interpreter's control state, so the
integer-stack fields are one part of a larger frame and can be exposed or
hidden on their own (\cref{tab:window-annotation-controls}).

\Cref{fig:push-pop} gathers these fields in one class and gives the ordinary
push and pop paths, abbreviated from the interpreter with \code{NTOP=2} and
\code{ACTIVE\_MAX=10}.  Both operations branch on \code{cache\_depth} and
compute frame indices from it, such as \code{si = dd - NTOP}.  When a trace
reads or writes a virtualizable array, RPython guards on the index value, so
each such index is a constant in that trace and the access becomes a trace
value instead of a load and a store.  Spill accesses, whose array is not
virtualizable, stay in the generated code.

\begin{figure}[!t]
\lstset{basicstyle=\scriptsize\ttfamily,aboveskip=0pt,belowskip=0pt,
        columns=fullflexible,keepspaces=true,language=Python,upquote=true,
        backgroundcolor=\color{figCode},
        keywordstyle=\color{black},
        stringstyle=\color{figEmph},
        emph={[1]t0,t1,cache_depth,frame,spill_ptr},
        emphstyle={[1]\color{figEmph}},
        emph={[2]spill},emphstyle={[2]\color{figAcc}\itshape}}
\begin{minipage}[t]{.30\linewidth}
\begin{lstlisting}
class IntStack(object):
  _virtualizable_ = [
    't0', 't1', 'cache_depth',
    'frame[*]', 'spill_ptr']

  def __init__(self):
    self.t0 = 0
    self.t1 = 0
    self.cache_depth = 0
    self.frame = [0] * NFRAME
    self.spill = [0] * NSPILL
    self.spill_ptr = 0
\end{lstlisting}
\end{minipage}\hfill
\begin{minipage}[t]{.32\linewidth}
\begin{lstlisting}
def push_on(self, v):
  dd = self.cache_depth
  if dd >= ACTIVE_MAX:
    self._push_on_full(v)
    return
  if dd >= NTOP:
    si = dd - NTOP
    self.frame[si] = self.t1
  self.t1 = self.t0
  self.t0 = v
  self.cache_depth = dd + 1
\end{lstlisting}
\end{minipage}\hfill
\begin{minipage}[t]{.36\linewidth}
\begin{lstlisting}
def pop_on(self):
  dd = self.cache_depth
  if dd <= 0:
    return self._pop_from_spill()
  r = self.t0
  self.t0 = self.t1
  if dd > NTOP:
    si = dd - NTOP - 1
    self.t1 = self.frame[si]
  self.cache_depth = dd - 1
  return r
\end{lstlisting}
\end{minipage}
\caption{The top fragment in the interpreter.  The scalar cells, the frame positions,
the occupancy, and the spill index are virtualizable (blue), whereas the
spill array is not (red italics).  Push and
pop select their path by testing \code{cache\_depth}; because \code{frame} is a
virtualizable array, RPython guards on the computed index \code{si}, which is
therefore constant inside the trace.}
\label{fig:push-pop}
\end{figure}

Two cases leave these paths.  When the top fragment is full, the push first moves the
deepest frame cell into the spill and shifts the remaining frame positions
before installing the new top.  When the top fragment is empty, the pop decrements
\code{spill\_ptr} and reads the cell at the resulting index, again without
building a callee frame.  An indexed access promotes the requested depth so that the
trace knows it, then selects \code{t0}, \code{t1}, a frame position, or a spill
offset, so a depth other than the recorded one fails that guard, and a hot
exit may get a bridge.

Normalization moves cells between the same fields and specializes in the same
way.  Once the occupancy and the target are known in a trace, its tests and
loop bounds are resolved and its frame indices are constants, and only its
stores of displaced cells to the spill remain.  An optional elision skips normalization when a
summary of the callee estimates that the top fragment cannot overflow during the
call.  The estimate is not a proof and can admit a call that spills; no
evaluated build enables elision.

\subparagraph*{Placing the floating-point top fragment.}
The floating-point top fragment mirrors the integer one, with fields \code{ft0},
\code{ft1}, \code{fframe}, \code{fdep}, and its own spill
(\cref{sec:architecture}).  For a primitive execution token, \code{CATCH}
snapshots this top fragment instead of saving only its depth.  We also built
the floating-point top fragment into the virtualizable frame.  On AWFY on Sapphire
Rapids, relative to a build whose floating-point stack is a contiguous array,
that configuration speeds up the three floating-point programs by
\(2.17\times\) (geometric mean) against \(2.13\times\) for the separate
object, but runs the eleven integer-only programs at \(0.97\times\) (down to
\(0.91\times\) on \code{Sieve}) against \(1.00\times\), so we kept the
separate object.  The two
configurations also differ in whether a call normalizes the floating-point
top fragment and whether its occupancy is promoted, so frame size alone may not
explain the gap.  Unlike the integer top fragment of RQ1, the floating-point top fragment
gains in RQ5 without being virtualizable.  We attribute this to promotion,
which no experiment isolates.  Its pop and push promote the occupancy
\code{fdep}, so every top-fragment position they touch is a constant in the trace, and the
optimizer's heap cache can track those loads and stores.  The unexposed integer
top fragment of RQ1 has no such promotion, and RPython guards its frame index only when
the array is virtualizable.
The controlled experiments of RQ1--RQ4 use the array build, in which the
floating-point array reference and depth are declared with the control state
(\cref{tab:window-annotation-controls}).

\begin{table}[tb]
\centering
\caption{The three builds compared in the Forth panels of \cref{fig:annotation-controls}.  They come from the earlier revision used for RQ1--RQ4, whose floating-point array has its reference and depth declared together with the control state, and differ only in which fields are declared virtualizable.  The evaluated build of \cref{sec:engines,sec:float-fragment} keeps the floating-point top fragment in a separate object instead (\cref{sec:architecture}).}
\label{tab:window-annotation-controls}
\small
\begin{tabular}{@{}lll@{}}
\toprule
configuration & integer top fragment & control state and floating-point array\\
\midrule
no annotations & off & off\\
integer stack unexposed & off & on\\
integer stack exposed & on & on\\
\bottomrule
\end{tabular}
\end{table}

\subparagraph*{What survives optimization.}
\Cref{fig:trace-example} shows the loop-carried part of two optimized traces
for the same small Forth word, with integer-stack annotations disabled or
enabled while retaining the other annotations.  In both traces, duplication
and arithmetic reduce to operations on SSA values: the unexposed configuration
does not reload the operand for every primitive.  The remaining difference
is synchronization with the stack representation.  Without integer-stack
annotations, each iteration still writes the scalar fields and a temporary
frame position.  With annotations, those writes disappear from the loop body.
The example illustrates access removal, not allocation removal or a timing
result.  Neither trace accesses the deep spill in this shallow loop.

\begin{figure}[tb]
\centering
\input{figures/forth/fig_forth_trace_example}
\caption{Loop bodies of two optimized traces of the same kernel, with
identical top-fragment shape and outer-interpreter rewrites, aligned row by row.  Both
compute the highlighted \code{dup dup * +} on SSA values (blue).  Without
integer-stack exposure, each iteration still writes the two scalar fields and
a frame position back (red, solid bar).  With exposure these writes disappear (dashed bar).
Modulo lowering, loop control, and long state-argument lists are abbreviated.}
\label{fig:trace-example}
\end{figure}

\subsection{Supported words and workload adaptations}
\label{sec:implementation-coverage}

\system{} implements all 133 Core words and 311 of the 371 words of the
standard word sets (\cref{tab:wordsets} in \cref{app:wordsets-appendix}),
covering what the measured workloads need; the benchmark ports keep a few
system-specific adaptations that \cref{sec:threats} reports.

The implementation reserves a spill of 16,384 cells under a ten-cell top fragment.
These are concrete capacities rather than an unbounded mathematical stack, and
the condition \(\dmax\le\bspill\) of \cref{sec:layout-correctness} describes
the reserved domain, not every state the implementation can reach near that
limit.  A push into a nonfull top fragment succeeds even when the spill is full, whereas
normalization and materialization raise a stack overflow when the spill lacks
room for the cells they move.

\subsection{Concrete costs}
\label{sec:costs}

\Cref{tab:costs} counts the work each operation does on the concrete
representation, taking word-sized arithmetic and one field or array access as
constant time and leaving cache behaviour aside.  Among the operations in \cref{tab:costs}, a full-fragment push, a normalization,
and a materialization can perform fragment-wide work, each in \(O(W)\) time.
The other listed operations perform a fixed number of field and array
accesses, whatever the stack's depth.

\begin{table}[tb]
\centering
\small
\caption{Interpreter costs.  The top fragment holds \(W=f+2\) cells, two scalar top
cells and a frame of width \(f\); \(U\) is the part of it that is occupied and
\(c\) the maximum number of cells a call-entry normalization keeps there.}
\label{tab:costs}
\begin{tabularx}{\linewidth}{@{}lXX@{}}
\toprule
operation & condition and concrete cost & worst case \\
\midrule
push & \(\Theta(1)\) if \(|U|<W\); \(\Theta(f)\) if full & \(O(W)\) \\
pop & \(\Theta(1)\), including an empty-fragment spill load & \(O(1)\) \\
depth-\(k\) read/write & \(\Theta(1)\): region tests and one field or array access & \(O(1)\) \\
call normalization & \(\Theta(|U|-c+\max(0,c-2))\) if \(|U|>c\); otherwise \(\Theta(1)\) & \(O(W)\) \\
return & \(\Theta(1)\), no data-stack transition & \(O(1)\) \\
materialization & \(|U|\) cell copies and \(\Theta(1+|U|)\) work & \(O(W)\) \\
\bottomrule
\end{tabularx}
\end{table}

The representation holds \(\Theta(\bspill+\wmax)\) cells, and each colon-word
\code{CATCH} record reserves \(\Theta(\wmax)\) more.  \system{} preallocates those records
up to its catch-depth limit, so \code{CATCH} copies into a reserved
record instead of allocating one.

\subparagraph*{Calls, exceptions, and materialization.}
Nested colon calls, entries from the host, and selected indirect and
exception-handling entries use call-entry normalization, whereas a tail call reuses the split as it is.  Ordinary call entry requires
no data-stack snapshot, so an indirect call needs neither a known target nor a
precomputed stack effect.  \code{CATCH} records the recovery state, and
a handled \code{THROW} restores the saved stack depths and pushes the
exception code, as the language requires~\cite[\S9.6.1.2275]{forth2012}.  The implementation restores the
saved top-fragment fields and spill index without undoing intervening writes to the
spill.  For a colon execution token, \code{CATCH} saves the scalar fields, the
occupancy, the spill index, and all \(f\) frame positions in a preallocated
record, whether or not they are occupied, at a cost of \(\Theta(\wmax)\).
\cref{sec:rq4} evaluates a variant that copies only the occupied frame
positions.  For a primitive execution token, \code{CATCH} instead allocates
separate top-fragment snapshots.
A \emph{materialization} empties the top fragment into the spill from the bottom
up, which restores one contiguous array at the price of one copy per occupied
cell.  In \system{} only the explicit word \code{STACK-MATERIALIZE}, which RQ4
uses, and an optional depth-only \code{CATCH} snapshot invoke it.  \code{SP@} returns the
address of a scratch cell holding a copy of the top value, so stable addresses
into the stack are outside what this implementation guarantees.

\section{Evaluation}
\label{sec:evaluation}

We evaluate \system{} in two parts.  First, we compare it as a complete system
with \gf{}, VFX Forth, and SwiftForth on three suites (\cref{sec:engines}).
These systems differ from \system{} in more than the top fragment, so this
comparison establishes competitiveness, not the top fragment's contribution.
Second, controlled experiments answer six questions:

\begin{description}
\item[RQ1] Does exposing the integer stack improve performance when the layout, call-entry policy, and other annotations are fixed?
\item[RQ2] How do traces differ between the top fragment and a flat cache that exposes the same ten cells?
\item[RQ3] How do the number of exposed cells, their shape, and call-entry policy affect performance?
\item[RQ4] Where does the top fragment's copying cost time?
\item[RQ5] Does the top fragment carry over to the floating-point stack without slowing integer-only programs?
\item[RQ6] Which observations transfer to a second interpreter?
\end{description}

\noindent RQ6 uses a second interpreter, \emph{RPyFactor}, which builds the same top fragment
into an RPython interpreter for a subset of Factor~\cite{factor10}
(\cref{sec:factor}).  Its results appear alongside the Forth ones from
\cref{sec:rq1-annotations} on.  Unless stated otherwise, \system{} uses the
stack fragment layout of \cref{sec:layouts} with \(f=8\) and \(c=2\).  Only
RQ3 and RQ6 test adaptive entry (\cref{sec:adaptive}).  Every policy
preserves the decoded stack (\cref{sec:layouts}), so all builds are
interchangeable.  The model predicts nothing about their performance.

\subsection{Experimental method}
\label{sec:method}

\subparagraph*{Programs.}
We use three suites, ordered by program size and reported separately.
\emph{Shootout} is 18 of the 25 kernels of the Shootout
collection~\cite{shootout}.  We omit \code{hello}, which is too short to time,
four kernels duplicated elsewhere (\code{composite}, \code{fibo},
\code{matrix}, \code{sieve}), and two whose results cannot be checked
(\code{moments}, \code{wordfreq}).  \emph{Are We Fast Yet} (AWFY) is our Forth
port of the 14 mid-sized object-oriented benchmarks of Marr et
al.~\cite{marr16}.  Three of them, \code{CD},
\code{Mandelbrot}, and \code{NBody}, use floating-point numbers.  The port
follows the Java version method by method.  An object is a zero-filled block
of cells, and a field is a word that adds an offset.  An overridden method is
an execution token stored in the object and called through \code{EXECUTE}.  A
closure is an execution token whose captured state lives in variables.
Allocation comes from an arena that each iteration resets, in place of a
garbage collector.  All 14 programs pass result verification on every system,
with VFX Forth and SwiftForth set to 53-bit x87 precision and SwiftForth given
a larger stack.  AWFY appears in \cref{sec:engines}, RQ1, RQ3, and RQ5.
\emph{Appbench} is five applications from the appbench suite used by Ertl and
Paysan~\cite{ertlpaysan24}: \code{benchgc}, \code{brainless}, \code{fcp},
\code{cd16sim}, and \code{lexex}, plus CoreMark~\cite{coremark}.

\subparagraph*{Machines, builds, and systems.}
All runs use Ubuntu 26.04 LTS with the performance governor on an AMD
Ryzen~9 5950X (Zen~3, SMT on) and an Intel Xeon w5-2455X (Sapphire Rapids,
SMT off).  Each \system{} binary is translated on its machine.  We never
average across machines.
\Cref{sec:engines,sec:float-fragment} use the build with top fragments on the data
and floating-point stacks (\cref{sec:window-implementation}).  RQ1--RQ4 use an
earlier revision whose floating-point stack is a contiguous array, with only
the array reference and depth virtualizable.  On the integer-only suites its suite
estimates are within about 1.5\% of the newer build's
(\cref{sec:float-fragment}).  There are three baselines.  \gf{}~0.7.9 is the
fast engine of the threaded-code interpreter Gforth with dynamic
superinstructions~\cite{ertlgregg03}.  We build it on each machine and
check that it uses them.\footnote{A \gf{} installation silently falls back to
plain threading when its dynamic code generation is unavailable, which would
understate it.}  VFX Forth~5.43 is a native compiler that inlines definitions and
optimizes data-stack traffic across them~\cite{vfxmanual}.  We run the same binary on both
machines.  SwiftForth~4.1.8 (4.1.10 on Sapphire Rapids) is a native compiler
with subroutine threading, inlining, and a rule-based optimizer~\cite{swiftmanual}.

\subparagraph*{Protocol.}
A process loads the program once and times 500 iterations.  Each
configuration of each program gets 100 processes pinned to one core, and the
configuration order rotates between repetitions.  A process contributes its
\emph{late-iteration median}, the median of its last 250 iterations.  In
\cref{sec:engines} it also contributes its first timed iteration, which runs
on a cold JIT.

Discarding early iterations is a warm-up heuristic, not a
steady-state test~\cite{kalibera13,barrett17}.  \Cref{app:rq3-appendix}
reports the drift between the two halves of the tail.  VFX Forth needs two
exceptions on Appbench.  It times in 10\,ms ticks, so a timed region holds 20
units of \code{brainless} and \code{cd16sim} and four of the others, and we
divide each region, the first included, by that count.  Its \code{lexex} cannot repeat in one
process, so each sample is a fresh process timing one unit.

\subparagraph*{Statistics.}
A program's value is the median of its process values.  A speedup is the
ratio of two such medians, baseline over treatment, and a suite estimate is
the geometric mean of the program ratios.  Iterations are therefore never
treated as independent samples.  Intervals are 90\% percentile intervals from
10,000 deterministic bootstrap resamples.  Each resample draws programs and
then process values within each program, and suites are resampled
independently.  We write an interval after its estimate, as in \(1.44\times\)
[1.25, 1.68], and do not read it as a significance
test~\cite{kalibera13}.  The layout-policy and RQ2 trace
analyses pair repetition indices, and the others draw each configuration's process
values independently.  In the speedup plots, each row shows every program as a small
point, a shaded band for the interquartile range with a tick at the median
(both computed on log ratios), and a large marker for the suite geometric mean
with its capped interval just outside it.  Where both machines appear, Zen~3
is the upper sub-row with a filled marker and Sapphire Rapids the lower one
with a hollow marker.

\subparagraph*{Workload characteristics.}
An instrumented run that contributes no timing counts four quantities per
program (\cref{fig:workloads}): the 95th-percentile integer-stack depth at a
push, word calls per 1,000 primitives, the share of stack accesses that reach
the spill, and the mean number of cells a normalization moves.  The Shootout kernels run in
a shallow stack and most rarely call words, with medians of 4 cells, 0.004 calls,
under 0.001\% spill accesses, and 0.04 cells per normalization.  AWFY and
Appbench are deeper and call-heavy.  Their medians are 20.5 and 14.5 cells,
32.4 and 22.7 calls, 5.0\% and 2.0\% spill accesses, and 0.41 and 0.60 cells
per normalization.  \Cref{sec:rq3} tests whether more cells matter more
there.

\subsection{Performance comparison with established Forth systems}
\label{sec:engines}

\Cref{fig:engines} compares \system{} with \gf{}, SwiftForth, and VFX Forth.
Pairs of numbers below are late-iteration speedups on Zen~3 and Sapphire
Rapids.

\begin{figure}[tb]
\centering
\input{figures/forth/fig_forth_engines}
\caption{Speedup of \system{} over three Forth systems per suite; values above
one favour \system{}.  \emph{late} uses the late-iteration median and
\emph{first} the first timed iteration, per unit where VFX Forth batches.
Rows follow the plot convention of \cref{sec:method}.}
\label{fig:engines}
\end{figure}

\subparagraph*{Shootout.}
\system{} is faster than all three systems.  It is \(2.04\times\) and
\(1.91\times\) faster than \gf{}, with 17 of 18 kernels faster on both
machines.  Against \swift{} it reaches \(4.60\times\) [2.58, 8.20] and
\(6.65\times\) [3.42, 12.82].  Against \vfx{} it reaches \(2.06\times\) [1.29,
3.38] and \(2.41\times\) [1.66, 3.51] and wins 13 and 15 kernels.  Two large
ratios, on \code{sumcol} and \code{spellcheck}, involve line input and the
adapted defining-word path of \cref{sec:threats}.

\subparagraph*{Are We Fast Yet.}
The mid-sized programs fall between the kernels and the applications.
\system{} is \(2.91\times\) [2.55, 3.34] and \(2.58\times\) [2.34, 2.82]
faster than \gf{} and \(2.07\times\) [1.78, 2.42] and \(2.48\times\) [2.07,
2.99] faster than \swift{}.  Against \vfx{} it is about as fast, \(1.07\times\)
[0.79, 1.48] and \(1.00\times\) [0.79, 1.29].  It wins the three
floating-point programs of \cref{sec:float-fragment}: \code{CD} by \(2.1\)--\(2.4\times\),
\code{Mandelbrot} by \(1.8\)--\(2.4\times\), and \code{NBody} by
\(3.5\)--\(4.0\times\).  It also wins \code{Storage} on both machines and
three more programs on Zen~3, and it loses most on
\code{Bounce}, \code{Permute}, \code{Richards}, and \code{Towers}, at
\(0.36\)--\(0.69\times\).

\subparagraph*{Appbench.}
\system{} is \(2.72\times\) and \(1.74\times\) faster than \gf{} and
\(1.78\times\) [1.31, 2.43] and \(2.04\times\) [1.46, 2.94] faster than
\swift{}.  It is slower than \vfx{}, at \(0.69\times\) [0.53, 0.87] and
\(0.78\times\) [0.56, 1.12], and the second interval includes one.  It
trails VFX Forth on five of the six applications, most on \code{cd16sim} at
about a third of its speed.  Its only win is \code{lexex}, for which VFX Forth
measures one sample per process.

\subparagraph*{First iteration.}
On a cold JIT, \system{} is slower than \vfx{} on all three suites, for
example \(0.64\times\) [0.46, 0.94] and \(0.56\times\) [0.43, 0.76] on AWFY.
Against \gf{}, it is already faster on AWFY, \(1.55\times\) and
\(1.39\times\).  Against \swift{}, its
first-iteration intervals include one on Shootout and Appbench.  Loading,
including the native compilers' compilation, is untimed for every system,
whereas JIT compilation inside a timed iteration is counted.  These numbers
therefore do not measure total compilation cost, startup, or break-even
time.

\subsection{RQ1: Does exposing the integer stack improve performance when the layout, call-entry policy, and other annotations are fixed?}
\label{sec:rq1-annotations}

The design combines two decisions.  The first moves the top cells out of the
stack array into the fields \code{t0}, \code{t1}, and \code{frame}.  The
second declares those fields virtualizable, so that the compiler carries them
as trace values instead of loading and storing them.  Classical stack caching
takes only the first decision~\cite{ertl95,ertl04,ertl05}, so we ask what the
second adds.  The cells could instead live in local variables of the dispatch
loop (RPython's red variables), which a trace also carries as values, but every
primitive that touches the stack would then have to receive and return them.
Virtualizable fields give the cells the same treatment while primitives keep
reaching the stack through the interpreter object.  We fix the two-scalar/eight-cell layout, the call-entry policy,
and every virtualizable field outside the integer stack, and vary only whether
the integer-stack fields are declared.  \Cref{sec:rq3} measures the call-entry
policy separately.  The three builds of \cref{tab:window-annotation-controls}
come from one source and run the same stack operations.  Both builds of the
integer-stack contrast have a virtualizable interpreter frame, so the contrast changes
which fields the frame exposes, not whether the mechanism is used.

\begin{figure}[!tb]
\centering
\input{figures/forth/fig_forth_annotation_controls}
\caption{What each group of annotations is worth, with the stack fragment layout and call-entry policy
fixed.  \emph{control/float only} adds the control and floating-point
annotations to an unannotated build, \emph{integer stack exposed} then adds
the integer-stack annotations, and \emph{all annotation groups} adds both to
the unannotated build; RPyFactor's \emph{control only} and \emph{data stack
exposed} play the same roles.  Values above one favour the build with more
annotations.  Rows follow \cref{sec:method}.  Forth uses 18 Shootout, 14 AWFY, and six
Appbench programs, Factor 20 kernels.}
\label{fig:annotation-controls}
\end{figure}

The effect is similar on the two machines (\cref{fig:annotation-controls}).
It is \(1.444\times\) [1.255, 1.693] and \(1.460\times\) [1.247, 1.719] on
Shootout, \(2.114\times\) [1.513, 3.183] and \(1.828\times\) [1.509,
2.267] on AWFY, and \(1.626\times\) [1.366, 1.964] and \(1.629\times\)
[1.394, 1.894] on Appbench (Zen~3/Sapphire Rapids), with all six intervals
above one.  On Zen~3, \code{Json} runs \(27\times\) faster and
the other 13 AWFY programs \(1.74\times\).  Exposing the control and floating-point
fields instead gives a smaller effect, and its intervals include one on
Shootout and on Zen~3 AWFY.  The two steps multiply to the
all-groups speedup, but integer-stack exposure alone is unmeasured.  The all-fields contrast also measures more than data-stack
visibility.


\Cref{fig:annotation-controls} also shows the same contrast in RPyFactor
(\cref{sec:factor}).  A separate Zen~3 comparison puts the annotated top-fragment build
at \(2.14\times\) [1.62, 2.84] on Shootout and \(2.80\times\) [2.22, 3.51] on
Appbench over an unannotated contiguous stack.  It changes representation
and annotation together, so it does not isolate visibility.

\subparagraph*{Answer to RQ1.}
With the stack fragment layout, the call-entry policy, and the other annotations fixed, exposing the
data stack raises all six Forth suite means and both Factor means, although
single programs such as \code{hash2} slow down.

\subsection{RQ2: How do traces differ between the top fragment and a flat cache that exposes the same ten cells?}
\label{sec:rq2-traces}

The sizes of traces and guard-exit records are an empirical question,
separate from the bound of \cref{lem:exit-bound}.  On all three suites, we compare the
default fragment layout with a conventional stack cache of the same size, the
\emph{flat ten-cell cache}.  It holds the top ten cells in ten scalar
fields and everything deeper in the ordinary stack array, and it skips call-entry
normalization.  Both layouts thus expose ten cells.  They differ in shape, one
flat set of fields against two scalars over a frame, and in call-entry policy, which
\cref{sec:rq3} separates.  We ran each layout in 100 fresh processes per
program and measured compilation time in a separate run without trace logging.
Over all optimized loops and bridges, we count operations, residual
spill-based array accesses, and values in guard-exit records.

On Zen~3, the top fragment emits fewer operations per optimized trace than the
flat ten-cell cache on 16 of the 18 Shootout kernels, 9 of the 14 AWFY
benchmarks, and all six Appbench applications.  Its guard-exit records hold
nearly as many values on Shootout and about 8\% and 7\% more on AWFY and
Appbench (\cref{app:rq2-appendix}).

\subparagraph*{Answer to RQ2.}
On Zen~3 with the same ten exposed cells, the top fragment emits 13\% fewer operations per trace
on Shootout and 25\% fewer on Appbench.  On AWFY the reduction is 6\%, with an
interval that includes no change.  On AWFY and Appbench it has more residual
spill accesses and slightly more guard-exit values.  Two kernels account for
most of the Shootout compile-time gain.

\subsection{RQ3: How do the number of exposed cells, their shape, and call-entry policy affect performance?}
\label{sec:rq3}

The top fragment differs from conventional top-of-stack caching in two ways.  It
exposes more cells, and it splits them into scalars, a frame, and a spill
instead of flat scalars over a spill.  Four builds separate these changes and
the call-entry policy, each changing one thing.

\begin{description}
\item[two-cell cache] two top values in scalar fields, everything deeper in one
  array.
\item[flat ten-cell cache] ten scalar positions instead of two, without
  call-entry normalization (\cref{sec:rq2-traces}).
\item[fragment/no-normalization] the same ten cells split two-plus-eight over the
  shared spill, still without normalization.
\item[fragment+normalization] that split, plus call-entry movement.
\end{description}

\noindent
All four preserve the decoded stack.  Read in order, they isolate the number of exposed cells,
then shape, then call-entry policy.

\begin{figure}[tb]
\centering
\input{figures/forth/fig_forth_stack_policy}
\caption{Controlled layout comparisons per suite.  All series are speedups
over the two-cell cache measured on the same machine, and comparing the two
top-fragment configurations isolates the incremental effect of call-entry
normalization.  Rows follow \cref{sec:method}, with 90\% paired two-level
bootstrap intervals.}
\label{fig:ablation}
\end{figure}

\subparagraph*{Number of exposed cells.}
Relative to the two-cell cache (\cref{fig:ablation}), the complete layout is 1.068 [0.993, 1.146] and 1.110
[1.010, 1.230] on Shootout, 1.246 [1.151, 1.346] and 1.179 [1.094, 1.277] on
AWFY, and 1.219 [1.095, 1.356] and 1.146 [1.058, 1.245] on Appbench
(Zen~3/Sapphire Rapids).  Widening the flat cache to ten positions exceeds
that on AWFY, at 1.301 [1.185, 1.419] and 1.192 [1.117, 1.283], and recovers
most of it on Appbench, at 1.182 [1.015, 1.357] and 1.102 [0.982, 1.220].  It
recovers less on Shootout, at 0.978 [0.889, 1.078] and 1.068 [0.947, 1.226].
The number of exposed cells alone can thus account for much of the gain on the larger programs.
The top fragment without normalization already reaches 1.332 and 1.239 on AWFY and
1.195 and 1.118 on Appbench, so the improvement there comes before any
call-entry movement.

\subparagraph*{Shape with ten exposed cells.}
Comparing the two ten-cell configurations without normalization holds both
the cell count and call-entry policy fixed.  The ratio of their suite estimates puts the
fragmented layout at 1.039 and 1.015 on Shootout, 1.023 and 1.039 on AWFY,
and 1.011 and 1.014 on Appbench (Zen~3/Sapphire Rapids); \cref{fig:ablation}
gives intervals only against the two-cell cache.
Against the flat ten-cell cache, the complete layout, which also differs in
call-entry policy, is 1.092 [1.028, 1.163] and 1.040 [0.967, 1.121] on Shootout,
0.958 [0.890, 1.034] and 0.989 [0.946, 1.047] on AWFY, and 1.031 [0.954,
1.113] and 1.040 [0.971, 1.112] on Appbench, with only the Zen~3 Shootout
interval excluding one.  On Sapphire Rapids, \code{objinst} holds the Shootout estimate down, and
removing it raises the estimate from 1.040 to 1.067.  On Zen~3, no single
program plays that role, and leave-one-out estimates span only 1.066 to
1.103.  There the result comes from programs that a flat cache handles badly:
\code{methcall}, \code{recurse}, \code{ack}, and \code{hash2} run at 0.66 to
0.81 of the two-cell cache when flat and at 0.90 to 1.21 with the complete layout.
Without matched access counts, these numbers do not explain why a given
program gains.

That pairing still mixes shape with call-entry policy, so we repeat it under a
matched adaptive top fragment, in which both use a per-word target \(c_w\) and
differ only in being fragmented or flat.  The shape effect is then 1.068
[1.026, 1.115] and 1.014 [0.956, 1.076] on Shootout, 1.092 [1.051, 1.133]
and 1.043 [1.005, 1.090] on AWFY, and 1.037 [0.967, 1.097] and 1.047 [1.019,
1.079] on Appbench (Zen~3/Sapphire Rapids).  All six intervals are
narrower than their fixed-entry counterparts.  Four exclude one: Zen~3 on Shootout,
both machines on AWFY, and Sapphire Rapids on Appbench.

\subparagraph*{Call-entry normalization.}
Normalization changes suite estimates little.  The ratio of the suite
estimates with and without it is 1.051 and 1.024 on Shootout, 0.936 and 0.952
on AWFY, and 1.020 and 1.025 on Appbench (Zen~3/Sapphire Rapids).  Of the six direct
paired intervals, only Zen~3 Shootout [1.003, 1.109] excludes one.  On Appbench it changes the spread instead.
Without the normalization, the worst Appbench program is
\code{fcp} at 0.849\(\times\) and 0.882\(\times\) of the two-cell cache.  With
it, the worst is \code{cd16sim} at 0.948\(\times\) and 0.979\(\times\).  It
improves \code{fcp} from 0.849 to 1.121 and from 0.882 to 1.111 but lowers
\code{lexex} from 1.379 to 1.135 and from 1.241 to 1.027.

Two further checks support the advantage of the complete layout over the
two-cell cache (\cref{app:rq3-appendix}).  Its point estimates stay above one
after screening all four main builds for drift and after omitting any one
program.  In a
parameter study over other splits, frame widths, and adaptive entry, no
setting dominates the default.

\subparagraph*{Answer to RQ3.}
The number of exposed cells accounts for much of the gain over a two-cell cache on the larger
programs.  Shape adds a small effect with the same ten cells.  Normalization
redistributes per-program results and moves suite estimates by at most about
6\%, with one of six intervals excluding one.  We keep the
two-plus-eight top fragment as a default, not a proven optimum.

\subsection{RQ4: Where does the top fragment's copying cost time?}
\label{sec:rq4}

The bounds of \cref{sec:costs} give the worst-case cost of the
representation.  This section asks where the fragment layout actually pays it.
We isolate three copying operations.  An exception snapshot saves the top fragment when
\code{CATCH} runs (\cref{sec:integration}).  A call-entry normalization moves
the cells below the top \(c\) into the spill (\cref{sec:layouts}).  A
materialization empties the top fragment into the array
(\cref{sec:integration}).

\Cref{app:rq4-appendix} measures each.  The exception-heavy kernel
\code{except} slows as the top fragment widens, to 0.33--0.35\(\times\) of the
default at a 64-cell frame.  Skipping inactive slots in its empty snapshots
speeds it up by up to \(4.46\times\) over fixed-slot copying.  The median
program moves an average of 0.04 cells per
normalization on Shootout, 0.41 on AWFY, and 0.60 on Appbench.  Forced
materialization,
timed as a whole operation, stays within 4\% of the unforced time at depths
32 and 128 by point estimate.  Full-fragment pushes also copy cells but are not
isolated here.

\subparagraph*{Answer to RQ4.}
Width costs most on the exception-heavy \code{except}, where an
occupancy-aware snapshot removes most of the penalty.  Normalization moves few
cells each time,
and forced materialization stays near the unforced time beyond the top fragment by
point estimate.  The
snapshot penalty comes from fixed-slot copying, not from the representation.

\subsection{RQ5: Does the top fragment carry over to the floating-point stack without slowing integer-only programs?}
\label{sec:float-fragment}

Forth's floating-point stack is a second call-shared stack.  We test whether
the top fragment carries over to it without slowing integer-only programs.  We
compare \rpyforth{} with both top fragments (\cref{sec:window-implementation}) with
three builds that replace one or both top fragments with a contiguous array whose
reference and depth are virtualizable. The
first, used in RQ1--RQ4, keeps the integer top fragment and uses a floating-point
array. The second uses an integer array and keeps the floating-point top fragment, and the
third uses arrays for both.  A ratio is the time of a reduced build over the
time of the build with both top fragments, so a value above one is what the missing
top fragments were worth.  All three suites follow the protocol of
\cref{sec:method}, with the four builds interleaved.

\begin{figure}[tb]
\centering
\input{figures/forth/fig_forth_frag_ablation}
\caption{What each top fragment is worth.  Each row is the time of a build with the
named top fragment(s) replaced by arrays, divided by the time of the build with both
top fragments; values above one favour keeping them.  The AWFY panel includes all 14
programs.  Rows follow
\cref{sec:method}.  \Cref{tab:frag-ablation} gives the Zen~3
values, including the floating-point and integer-only AWFY subsets.}
\label{fig:frag-ablation}
\end{figure}

The two top fragments serve different programs (\cref{fig:frag-ablation}).
Removing the floating-point top fragment keeps the integer-only suite estimates near
one: \(1.01\times\)
[0.98, 1.03] and \(1.00\times\) [0.99, 1.01] on Shootout, \(1.01\times\)
[1.00, 1.02] and \(1.00\times\) [1.00, 1.01] on the eleven integer-only AWFY
programs, and \(0.99\times\) [0.96, 1.01] and \(1.00\times\) [0.99, 1.02] on
Appbench (Zen~3/Sapphire Rapids).  On the three floating-point programs it
costs \(2.31\times\) [1.63, 3.27] and \(2.13\times\) [1.63, 2.78], from about
\(1.4\times\) on \code{CD} to \(3.2\)--\(3.9\times\) on \code{Mandelbrot}.
Removing the integer top fragment costs \(1.83\)--\(1.91\times\) on Shootout,
\(2.35\)--\(2.60\times\) on the integer-only AWFY programs, and
\(2.16\)--\(2.46\times\) on Appbench.  Unlike RQ1, this contrast changes
representation and annotation together.  On the integer-only suites, removing
both top fragments costs about as much as removing the integer top fragment alone.  The
floating-point programs instead lose \(3.37\times\) [2.85, 3.98] and
\(3.01\times\) [2.48, 3.66], more than either top fragment alone costs them
(\(1.79\)--\(2.31\times\)), so they benefit from both.  Repeating one Shootout build
against itself gives a median ratio of 1.007 and a maximum of 1.217 on Zen~3,
so we do not interpret single-kernel differences below about \(1.2\times\).

\subparagraph*{Answer to RQ5.}
The top fragment carries over to the floating-point stack.  Placed outside the
frame that integer-only traces carry, it speeds up the programs that use that
stack and leaves the integer-only suite estimates within about 1.5\%.

\subsection{RQ6: Which observations transfer to a second interpreter?}
\label{sec:factor}

RQ6 asks which observations transfer to a second interpreter.  RPyFactor is an
RPython interpreter for a subset of Factor~\cite{factor10}, a concatenative,
quotation-based language that also shares its data stack across calls. RPyFactor's
adaptive selector uses the classes and long-body allowance of
\cref{sec:adaptive} but is quotation-aware.  Its scan follows literal
quotations into combinators such as \code{dip}, \code{call}, and \code{if},
taking the deeper branch of an \code{if}, and reuses summaries of scanned
words.  It stops after a looping combinator or at an unknown word and
keeps the depth found so far, so it needs a partial body summary, not
a full stack effect.

Seven builds share one interpreter source. The \emph{boxed-list baseline}
keeps the stack in an unbounded boxed list. \emph{Fixed entry (2),
unannotated} adds the top fragment but hides it from the optimizer, and \emph{fixed
entry (2)} makes the same top fragment virtualizable. \emph{No entry normalization}
keeps that top fragment but drops the call-entry movement.  \emph{Fixed entry (4)}
and \emph{fixed entry (8)} set \(c\) to four and eight.  \emph{Adaptive
entry} sets a per-definition target from a static scan that follows
the literal quotations of \code{if}, \code{while}, and \code{times}.  Native
Factor, timed in-process with \code{tools.time}, is the outside reference.

The benchmark suite is 20 ported kernels: four from Shootout~\cite{shootout},
ten from the Forth list and recursion benchmark set, and six from Factor's own
benchmark vocabularies~\cite{factorbench} (\code{nsieve-bytes},
\code{binary-trees}, \code{heaps}, \code{lcs}, \code{euler150},
\code{fannkuch}).  These six run inner loops 6 to 20 cells deep, and the rest stay shallow.  Factor has no application suite comparable to
Appbench, so this comparison covers kernels only.  The annotation contrast
runs on both machines, and the seven-build policy comparison and the native
reference run on Zen~3, all under the protocol of \cref{sec:method}.

\subparagraph*{Exposing the data stack.}
As in Forth, the control configuration keeps \code{cs\_ptr} virtualizable,
and only the data-stack fields are toggled (\cref{fig:annotation-controls}).
Exposing them gives \(1.421\times\) [1.246, 1.631] on Zen~3 and
\(1.556\times\) [1.404, 1.717] on Sapphire Rapids.  The gain is not universal.
On Zen~3, \code{nrev} (0.767), \code{sum} (0.820), \code{step} (0.807), and
\code{filter} (0.962) fall below one.  On Sapphire Rapids, \code{fibrec} is
0.949 and \code{random} is unchanged.  We discard no samples.  The four Zen~3
regressions drift by more than 10\% between the halves of their tails.  On
Sapphire Rapids, every curve that crosses this threshold comes from
\code{mapfold}.

\begin{figure}[tb]
\centering
\input{figures/forth/fig_forth_factor}
\caption{RPyFactor speedups on 20 ported programs, Zen~3.  Left:
representation and frame-virtualization settings relative to the boxed-list
baseline.  Right: call-entry policies relative to the same virtualizable
top fragment without normalization, so \emph{fixed entry (2)} appears in both panels
against different baselines.  Rows follow \cref{sec:method}.}
\label{fig:factor}
\end{figure}

Against the boxed-list baseline, fixed entry~(2) gives \(1.559\times\)
[1.388, 1.723] unannotated and \(2.289\times\) [1.992, 2.614] virtualizable
(\cref{fig:factor}).  In aggregate, every fixed call width and adaptive entry
runs at 0.98--1.00 of the unnormalized top fragment, and all intervals include
one.  Per program, however, the
choice matters and no policy dominates.  Adaptive entry improves \code{fact}
(\(1.39\times\)), \code{tak} (\(1.16\times\)), \code{binary-trees}
(\(1.13\times\)), and \code{fibrec} (\(1.13\times\)).  It loses on
\code{heaps} (\(0.72\times\)), \code{fannkuch} (\(0.81\times\)), and
\code{euler150} (\(0.87\times\)), and leaves the rest within 0.97--1.02.  These
three programs share an inner-loop shape.
Mid-body, they reread cells below the top that local pushes have carried past
any entry target, so normalization may move those cells into the spill.
Testing this explanation would need matched counts of spill accesses.  This cost
differs from the fixed-width snapshot copying of Forth's \code{except}.

Against native Factor's optimizing compiler, the geometric-mean ratio of
native time to adaptive entry is \(1.24\times\) [0.89, 1.80].  RPyFactor is
faster on \code{nestedloop} (\(20.6\times\)) and \code{random}
(\(8.7\times\)) and slower on \code{euler150} (\(0.37\times\)), \code{fibrec}
(\(0.46\times\)), and \code{heaps} (\(0.52\times\)).  The interval includes
one and per-program variation is large, so RPyFactor is merely competitive
here.

Trace counts compare
call-entry policies within one shape, not Forth's fragmented and flat layouts.  Per
optimized trace, the boxed-list baseline carries 2.8\(\times\) the operations
and 2.0\(\times\) the guards of the unnormalized top fragment, with 1.10\(\times\)
its compile time.  Normalization then adds work rather than removing it:
unnormalized over normalized per-trace counts are \(0.85\) for operations and
\(0.90\) for guards at \(w=2\), and \(0.88\) and \(0.91\) for adaptive
entry.  The
added work concentrates in the recursive and deep-stack programs above, while
compile time falls by about 10\% under both.  Frequent quotation calls are a
plausible source, but these counts do not isolate normalization traffic.

\subparagraph*{Answer to RQ6.}
The top fragment is practical in a second interpreter, and the exposure benefit
reproduces there on both machines.  No call-entry policy dominates, and adaptive
entry is about as fast as the unnormalized top fragment.  We did not measure a flat ten-cell Factor
build, so the shape contrast is untested outside Forth.

\section{Discussion}
\label{sec:discussion}

\subsection{What to expose before choosing a policy}
\label{sec:generality}
\label{sec:policy-discussion}

Preserve the shared stack sequence across calls and expose a bounded set of
individually identifiable locations to the optimizer.  With the stack fragment layout and call-entry
policy fixed, exposing those locations improves suite-level performance in
both realizations on both machines
(\cref{sec:rq1-annotations,sec:factor}).  Shape and call-entry policy matter less, and no adaptive policy dominates the fixed
default.  Against the two-cell baseline, a flat ten-cell cache has
higher point estimates than the top fragment on AWFY and recovers most of its gain on
Appbench.  Fixed normalization lowers AWFY point estimates by 5--6\%, but the
paired intervals include one.  Limited effort
should go to the interface, not to a selector.

Boundary operations deserve separate treatment.  Empty exception snapshots need not pay for unused frame capacity, as the
occupancy-aware snapshot on \code{except} shows.  Frequent normalization can
add spill traffic, but it moves few cells per call on average
(\cref{sec:rq4}).  Neither is a reason to abandon a bounded representation.

\subsection{What adopting the top fragment requires}
\label{sec:adoption}

The top fragment is cheap to declare and not free to maintain.  An implementer takes
on four obligations, all of them visible in \cref{sec:implementation}.  The
fill order must hold, with scalar positions occupied before frame positions, or
the decomposition of a top-fragment sequence stops being unique and the simulation
of \cref{lem:three-region} no longer applies.  Operations that need a contiguous stack must materialize any occupied top fragment.
In \system{}, only explicit requests and an optional depth-only snapshot do so.
Stable stack addresses, as \code{SP@} could provide, would also need an
aliasing policy for later movement.  Exception recovery must preserve the top fragment.  Fixed-slot \code{CATCH} snapshots
of colon words copy every frame slot, so their cost grows with width
(\cref{sec:rq4}).  Finally every hot top-fragment access must reach a position the trace can
name: the integer top fragment gets constant frame indices because RPython guards on
the index of a virtualizable array access, and the floating-point one
promotes its occupancy.

None of this requires a static stack-effect analysis.  The refinement argument covers the fill
order for the modeled operations, RQ4 measures explicit materialization and one
exception-recovery strategy, and RQ1 measures exposure with constant
positions in place.  An adopter must still check that the spill, under its overflow
policy, holds what normalization and materialization move into it.

\subsection{What the host compiler could offer}
\label{sec:host-implications}

One observation points at the meta-tracing framework rather than at the VM.
With the same ten exposed cells, guard-exit records carry about as many values on Shootout
and about 8\% and 7\% more on AWFY and Appbench (\cref{sec:rq2-traces}).  The comparison also
changes call-entry policy, so it does not show why.  If a virtualizable frame is
serialized by field, recording inactive slots could contribute.  An interface
that let the VM declare which positions are occupied, as the occupancy field
already tells the trace, could let guard-exit records and the exception
snapshot of \cref{sec:rq4} omit inactive positions.  We do not measure
that benefit.

The remaining Appbench gap to VFX Forth is the other open item.  Compared with
Shootout, Appbench has deeper stacks at pushes and more word calls per
primitive (\cref{fig:workloads}), so spill traffic and call boundaries are
natural suspects.  The cross-system timings cannot separate them from the rest
of a mature native compiler (\cref{sec:engines}).

\subsection{Threats to validity}
\label{sec:threats}

\subparagraph*{What the controls do not isolate.}
The contiguous baseline of \cref{sec:rq1-annotations} changes representation
and annotation together, and we never measure a build that declares every
array position, so the cost of exposing a whole array is untested.  The rejected
in-frame placement of the floating-point top fragment rests on a preliminary
measurement without intervals that also varies normalization and promotion.  No
build varies only the promotion of the floating-point occupancy, so its role in
the floating-point gain is explained, not measured.

\subparagraph*{How far the results reach.}
\label{sec:other-vms}
The decoding argument covers the abstract operations of \cref{app:layout} and
predicts nothing about optimizer behavior.  Both machines share an instruction
set, and both realizations use RPython, so other instruction sets and other
JIT frameworks, such as Truffle and yk, remain untested.  yk meta-traces C
interpreters and offers promotion and idempotent functions but no counterpart
to virtualizables~\cite{yk}, so a top fragment there would have to live in
locals.  Other concatenative
languages motivate further realizations~\cite{factor10,joymanual,plrm3,herzberg09},
but each would need its stack operations, recovery rules, aliasing semantics,
and the host JIT's state interface checked.  Logical operations such as
\code{DEPTH}, \code{PICK}, and \code{ROLL} work across the fragment/spill split,
but stable pointers into the data stack do not (\cref{sec:adoption}).  The Forth contiguous baseline and the RPyFactor policy, native-Factor, and
trace-metric comparisons run on Zen~3 only.  RPyFactor implements a subset of
Factor with 20 kernel-sized programs.  Five of the six ports from Factor's
benchmarks are size-reduced, and \code{primrec} and \code{binrec} are builtins
in RPyFactor but prelude definitions in native Factor.

\subparagraph*{What the cross-system numbers can carry.}
The commercial engines run compatibility code for \code{NEXTNAME} in
\code{hash}, \code{hash2}, and \code{spellcheck} that redefines \code{CREATE}
and \code{CONSTANT}, and VFX \code{lexex} is timed one unit per process
(\cref{sec:engines}).  These timings compare complete engine configurations.  Controlled contrasts
(\cref{sec:rq1-annotations,sec:rq3}) estimate the effect of each design choice.  The Are We Fast Yet comparison rests on our own Forth ports and on the
x87 precision settings that VFX Forth and SwiftForth need to reproduce the
floating-point results.  The ports differ from the Java originals: allocation
comes from an arena reset every iteration, so benchmark objects are never
garbage collected, dynamic dispatch is an explicit \code{EXECUTE} through a stored
execution token, and integers are 64-bit cells, with 32-bit wrap-around masked
where the originals rely on it.  We compare Forth systems on the same ports
only.

\section{Related Work}
\label{sec:related}

\subparagraph*{Position.}
This work connects interpreter stack caching~\cite{ertl95,ertl04,ertl05}
with interpreter-state virtualization~\cite{bolz11,rpythonvirtualizables,%
truffleframe,truffleframedescriptor}.  Virtualization mechanisms are not
intrinsically restricted to call-local state.  We supply them a bounded,
decodable representation of a call-shared stack.

\subparagraph*{Forth stack caching.}
Ertl analysed static and dynamic stack caching for interpreters~\cite{ertl95}.
Gforth combines stack caching~\cite{ertl04,ertl05} with dynamic
superinstructions and other interpreter optimizations~\cite{ertlgregg03,ertl04,vmgen02}.  Our top fragment retains top-of-stack caching and exposes a fixed set of slots to the
tracer without precomputed stack effects or cache-state-specific handlers.  RQ1
isolates exposure, RQ2 compares traces against a flat ten-cell
cache, and RQ3 separates cell count, shape, and call-entry policy.

\subparagraph*{Native Forth compilation.}
Native compilers infer stack effects over regions and assign stack values to
machine registers~\cite{ertl96,vfxmanual,swiftmanual}.  Our representation
instead exposes a bounded top fragment to an existing meta-tracing JIT compiler
without static stack-effect analysis.  Ertl and Paysan compare gforth with
SwiftForth, VFX Forth, and gcc and trace a SwiftForth slowdown on
\code{cd16sim} to microarchitectural pitfalls rather than to the instructions
it executes~\cite{ertlpaysan24}.  Cross-system timings thus mix many factors,
so we use VFX Forth and SwiftForth only as external references and compare
layouts within one interpreter.

\subparagraph*{Interpreter-state optimization in meta-JIT frameworks.}
RPython builds a meta-tracing JIT compiler from an
interpreter~\cite{bolz09,bolztratt15}.  Its allocation removal keeps objects
allocated within a trace as trace values~\cite{bolz11}, and virtualizables
extend this to long-lived interpreter state such as
frames~\cite{rpythonvirtualizables}.  PyPy declares each Python frame
virtualizable, including its value stack as a fixed-size
array\footnote{\url{https://github.com/pypy/pypy/blob/51a08c6f0c21ef6a9d7a794db6d3f76fb7b0b558/pypy/module/pypyjit/interp_jit.py}}, but a call-shared Forth stack has no per-frame bound to
size such an array.  Gai{\ss}ert et al.\ build an RPython interpreter for
effect handlers, including Koka, whose call stack, the \emph{metastack}, is a
linked list of immutable stack segments; it relies on allocation removal to
keep frames and segments out of traces~\cite{gaissert25}.  Truffle instead
compiles self-optimizing AST interpreters~\cite{wurthinger12} by partial
evaluation, whose escape analysis and scalar
replacement~\cite{stadler14,wurthinger17} remove a \code{VirtualFrame} whose
slots a \code{FrameDescriptor} lays out~\cite{truffleframe,truffleframedescriptor}.
Larose et al.\ compare AST and bytecode interpreters on both frameworks and
find them about equally fast once compiled~\cite{larose23}, whereas we vary
one design choice, the stack representation, under a fixed compiler.
Unlike PyPy and Truffle, which size state per frame, and unlike the
metastack, which relies on allocation removal, our stack fragment layout gives
RPython's existing virtualizable mechanism a bounded, decodable representation
of a call-shared stack.  Our contribution is this representation, not a new
optimizer.

\subparagraph*{Call-shared stacks in other VMs.}
Several VMs share stack storage across calls while providing static information
about its use.  Lua~5.0 passes arguments in the caller's free registers, and
these registers become part of the callee's activation record.  Each record has
room for all of its function's registers~\cite{lua05}.  An in-place
WebAssembly interpreter makes the arguments the callee's first locals on the
caller's operand stack, and validation models that stack through each function
body~\cite{titzer22}.  Factor enforces a declared stack effect on every
ordinary word definition~\cite{factor10}.  Forth, PostScript, and Joy require neither declared
frames nor declared stack effects~\cite{forth2012,plrm3,joymanual}.  Our stack
fragment layout requires none of this information, so it applies to these
languages as well as to the VMs above.

\subparagraph*{Factor compilation.}
Factor's optimizing compiler checks stack effects by abstract interpretation
and builds its high-level IR from that analysis~\cite{factor10}.  It therefore already
resolves the source-language stack into SSA values at compile time.  Our design instead supplies a
bounded state representation to an existing meta-tracing JIT compiler without
requiring a per-word stack-effect solution.  We carry the representation into
RPyFactor to test whether it depends on Forth, not to compete with Factor's
compiler.  Its subset and ported benchmarks limit that experiment.

\subparagraph*{Register windows and stack machines.}
RISC register windows overlap for parameters, a caller's outs being the
callee's ins, and give each window private locals.  A return switches to the activation's saved
window~\cite{patterson81,sparcv9}.  The \layout{} instead follows logical stack
depth and has no activation identity or saved window.  Hardware stack machines
favor fast top-of-stack access~\cite{koopman89}, and the JOP Java processor keeps
each method's locals and saved caller context as a frame on its operand
stack~\cite{schoeberl05}.  Our top fragment shares one stack across
calls, without private operand frames.

\subparagraph*{Interpreter optimization.}
Threaded code, superinstructions, replication, and generated specialized
interpreters reduce dispatch
overhead~\cite{bell73,deutsch84,ertlgregg03,shi08,deegen}.
These are complementary: our motivating failure occurs after dispatch is gone,
when the residual program still manipulates the data-stack representation.

\section{Conclusion}
\label{sec:conclusion}

A meta-tracing JIT compiler such as RPython's lets an interpreter expose its
state as a fixed set of named slots.  A data stack that word calls share has
no frame boundary known before execution and hence no fixed per-call slots.
This paper keeps the top cells of such a stack in a fixed-width top fragment of
named slots, keeps every deeper cell in a shared spill array, and joins the two
with a decoding function into the logical stack.  Within the modeled capacity,
every modeled stack operation on this layout decodes to the same operation on
the logical stack, and every call-entry policy preserves the decoded sequence
(\cref{thm:refinement}).  Under \cref{lem:exit-bound}'s assumptions, a trace
exit rebuilds the data stack from at most \(\wmax\) top-fragment values and
\(O(1)\) control values.  Adopting the top fragment requires a fixed fill order,
materialization before contiguous access, exception recovery that
preserves it, and constant positions in hot traces
(\cref{sec:adoption}), but no static stack-effect analysis.

With the stack fragment layout, the call-entry policy, and the other
annotations fixed, exposing the integer stack gives geometric-mean speedups of \(1.44\)--\(1.46\times\) on 18
Shootout kernels, \(1.83\)--\(2.11\times\) on 14 AWFY benchmarks, and about
\(1.63\times\) on six Appbench applications in \system{}, with ranges across
two machines.  In RPyFactor, exposing the stack gives
\(1.42\)--\(1.56\times\) on 20 ported kernels.  With the same ten exposed cells, the
top fragment emits 13\% and 25\% fewer operations per trace than a flat cache on
Shootout and Appbench, and
call-entry normalization moves suite means by at most about 6\%, with the
best policy varying by program.  At late iterations, \system{} is
\(1.74\)--\(2.91\times\) faster than \gf{} and \(1.78\)--\(6.65\times\)
faster than SwiftForth on all three suites.  Against \vfx{}, it is
\(2.06\)--\(2.41\times\) faster on Shootout, \(1.00\)--\(1.07\times\) on
AWFY, and \(0.69\)--\(0.78\times\) on Appbench, where VFX Forth is faster.
Placing a second top fragment over the floating-point stack outside the
virtualizable frame, instead of a contiguous array, gives a
\(2.1\)--\(2.3\times\) geometric-mean speedup on the three floating-point
AWFY programs and changes integer-only suite means by at most about 1.5\%.

\subparagraph*{Future work.}
Four directions follow from this work.  First, its findings could move into
RPython itself.  Virtualizables expose fixed fields and whole arrays.  A
framework-level stack declaration with a bounded, occupancy-tracked region
could manage the top fragment, the spill, and trace-exit reconstruction, so that a
VM author declares a stack instead of building one.  Second, the layout could
be carried to other interpreter frameworks.  Truffle carries interpreter state
in frames under partial evaluation~\cite{truffleframe,truffleframedescriptor},
and yk offers no counterpart to virtualizables~\cite{yk}, so
each needs its own way to expose a bounded part of a call-shared stack.
Third, the top fragment could serve other stack-based VMs.  PostScript and Joy share
their stacks under semantics outside Forth's core, such as PostScript's
dictionary stack~\cite{plrm3} and Joy's quotation-based control~\cite{joymanual}.  Fourth, the findings may
carry to VMs with fixed-size frames.  Such a frame sizes its operand stack for
the deepest point of a function, but usually only part of it is occupied.
Exposing only the occupied top might shorten traces there too.

\subparagraph*{AI Usage}
Parts of the implementation, of the benchmarking and
plotting scripts, and of the prose in this paper were produced with the
assistance of generative AI tools (Anthropic Claude, used through Claude
Code) for coding support and editing.  The authors verified, tested, and
revised all such output, and take full responsibility for the content of
this paper, including its correctness and its experimental results.

\bibliography{references}

\appendix
\ifsupplemental
  \crefalias{section}{appendix}
  \crefalias{subsection}{appendix}
  \crefalias{subsubsection}{appendix}
\fi

\section{The Fixed-Width Top Fragment}
\label{app:layout}
This section defines the \layout{} and the assumptions the proofs rest on.

The interpreter object and the stack array both live as long as the VM, so no
allocation is ever removed here and the opportunity is the removal of accesses.
A location the optimizer can identify individually can be carried through
optimized code as a value and written back on exit, by frame virtualization or
scalar replacement.  A location selected by a run-time index into a large array
cannot: the optimizer must leave the load or store in place, specialize the
index, or model the array's elements as further state.

The reconstruction bound (\cref{lem:exit-bound}) therefore assumes a fixed number of individually identifiable top fragment
locations, a shared spill whose deep cells are in memory at an exit and whose
position is given by a constant number of references and indices, and constant
metadata per exposed location.  These are sufficient assumptions about the
representation, not necessary conditions for optimization, and existing frame
virtualization already offers this interface without demanding that every value
be absent from memory~\cite{bolz11,rpythonvirtualizables,truffleframedescriptor,%
truffleframe,wurthinger17}.

One contiguous array meets the language semantics but supplies no bounded
top fragment: a depth-\(k\) access computes its index from the stack pointer, so the
trace records an array operation, and declaring the whole array only trades
residual accesses for optimizer state.  Splitting the array per call would
bound each part but change the semantics, since a private frame that protects
or restores cells is a different data stack.

\subsection{Representation and decoding}
\label{app:representation}

The model hides the implementation's scalar and frame regions and treats every
cell outside the spill as one sequence.  A physical state is
\(q=\langle\spillseq,\winseq\rangle\), where both components are finite
sequences ordered from bottom to top: \(\spillseq\) is the occupied prefix of a
backing array of capacity \(\bspill\), and the cells in \(\winseq\) occupy
locations the optimizer can identify individually.  The parameters satisfy
\(1\le\wmax\) and \(0\le\dmax\le\bspill\), where \(\wmax\) is the maximum
number of top cells held outside the spill array and \(\dmax\) is the logical
data-stack capacity.  A state is valid, \(\operatorname{Inv}(q)\), when
\(|\spillseq|+|\winseq|\le\dmax\), \(|\winseq|\le\wmax\), and
\(|\spillseq|\le\bspill\); its logical stack is
\(\decode(q)=\spillseq\mathbin{\|}\winseq\), with \(\mathbin{\|}\) sequence
concatenation.  The condition \(\dmax\le\bspill\) reserves enough backing
capacity for the entire logical stack, so moving top-fragment cells into the spill
cannot introduce an overflow the abstract Forth stack would not observe.  These sequences describe semantics only, and \cref{sec:costs} bounds the
concrete representation.  The domain is capacity-reserved: it excludes the states of total
depth above \(\bspill\) that the implementation can reach near its limit.

\subsection{Stack operations}
\label{app:operations}

The transition rules operate on \(\spillseq\) and \(\winseq\) and state their
capacity and underflow conditions explicitly.  Pushing \(x\) reports overflow
when \(|\spillseq|+|\winseq|=\dmax\).  Otherwise, when \(|\winseq|<\wmax\), it
produces
\(\langle\spillseq,\winseq\mathbin{\|}[x]\rangle\).  When the top fragment is full,
write \(\winseq=[y]\mathbin{\|}\mathit{rest}\), where \(y\) is its deepest
cell; the result is then
\(\langle\spillseq\mathbin{\|}[y],\mathit{rest}\mathbin{\|}[x]\rangle\), and
the capacity invariant guarantees room for \(y\) in the spill.  Pop requires
\(|\spillseq|+|\winseq|>0\).  If \(\winseq=\mathit{rest}\mathbin{\|}[x]\), it
returns \(x\) and leaves \(\langle\spillseq,\mathit{rest}\rangle\); otherwise,
writing \(\spillseq=\mathit{rest}\mathbin{\|}[x]\), it returns \(x\) and leaves
\(\langle\mathit{rest},[]\rangle\).  An empty top fragment is therefore not a stack
underflow while the spill is nonempty.  For
\(0\le k<|\spillseq|+|\winseq|\), a depth-\(k\) read selects
\(\winseq[|\winseq|-1-k]\) when \(k<|\winseq|\) and
\(\spillseq[|\spillseq|-1-(k-|\winseq|)]\) otherwise; a write replaces the cell
selected by the same cases.  Constant depths in common Forth words let the
tracer fold the tier selection after observing a stable top-fragment occupancy; the
abstract rule does not prescribe how a changing depth is implemented, and
\cref{sec:rpython} describes the guarded specialization \system{} uses.  Depth
returns \(|\spillseq|+|\winseq|\) and reset produces \(\langle[],[]\rangle\);
neither depends on a call frame.

\subsection{Calls and materialization}
\label{app:calls}

A control-only call or return acts as the identity on the stack state.  The representation additionally permits an optional call-entry
normal form: for \(0\le\ccall\le\wmax\), uniquely split
\(\winseq=\mathit{moved}\mathbin{\|}\mathit{kept}\) with
\(|\mathit{kept}|=\min(\ccall,|\winseq|)\), and normalization produces
\(\langle\spillseq\mathbin{\|}\mathit{moved},\mathit{kept}\rangle\).  It
preserves total depth and reduces the occupancy bound from
\(\min(\wmax,\dmax)\) to \(\min(\ccall,\dmax)\).  It does not make the data stack call-local: a
callee pop that exhausts \(\mathit{kept}\) continues with the cells just
appended to \(\spillseq\), and return leaves the state unchanged.  Tail-call
transfers skip it, and the implementation uses \(\ccall=2\) by default.

Materialization is the transition
\(\langle\spillseq,\winseq\rangle\longrightarrow
\langle\spillseq\mathbin{\|}\winseq,[]\rangle\), whose result the capacity
invariant guarantees fits.  The operations evaluated here do not expose the
physical address of a stack slot; a word returning stable addresses into a
contiguous data-stack array would require materialization and an aliasing
policy, outside the present implementation claim.

\subsection{Refinement to one abstract stack}
\label{app:refinement}

The policy that moves cells between top fragment and spill is a free parameter, so
the obligation is to show that the split changes neither cell order, nor
returned values, nor what a callee sees.  Guard recovery is a separate matter:
it relies on the host JIT rebuilding interpreter state correctly, which this
model does not cover.

An abstract Forth stack is one sequence \(A\) ordered from bottom to top with
\(|A|\le\dmax\).  Push appends a value when \(|A|<\dmax\), and pop removes the
last one when \(A\) is nonempty.  A depth-\(k\) read or write with
\(0\le k<|A|\) accesses \(A[|A|-1-k]\), depth returns \(|A|\), and reset
empties \(A\).  Decoding erases the physical split, so it suffices to
show that each operation has one of these effects after decoding, which is what
lets the result quantify over policies.  A
\emph{top-fragment policy step} is any state change
\(\langle\spillseq,\winseq\rangle\rightarrow
\langle\spillseq',\winseq'\rangle\) with
\(\spillseq'\mathbin{\|}\winseq'=\spillseq\mathbin{\|}\winseq\) and
\(|\winseq'|\le\wmax\).  Call normalization for any \(\ccall\),
materialization, the identity step of an ordinary call, and a policy that
refills the top fragment from the spill are all instances.

\begin{restatable}[Invariant preservation]{lemma}{lemInvariant}
\label{lem:invariant}
If \(\operatorname{Inv}(q)\) holds, every successful operation that changes
\(q\) produces another state that satisfies the invariant.  Reads, depth,
and control-only calls and returns leave \(q\) unchanged.
\end{restatable}

Proof sketch: push is allowed only below \(\dmax\) and moves at most one cell
into the spill; pop, reset, and indexed write only shrink or preserve lengths;
a top-fragment policy step preserves the total length and respects \(\wmax\) by
definition, and \(\dmax\le\bspill\) leaves room for whatever it moves into the
spill.

\thmRefinement*

Proof sketch: decoding concatenates the spill and the top fragment, so each case
reduces to an identity on sequences.  A push with room appends \(x\) to
\(\winseq\) and hence to the decoded stack.  A full-fragment push with
\(\winseq=[y]\mathbin{\|}\mathit{rest}\) yields
\(\spillseq\mathbin{\|}[y]\mathbin{\|}\mathit{rest}\mathbin{\|}[x]
=\spillseq\mathbin{\|}\winseq\mathbin{\|}[x]\).  Pop removes the last cell of
\(\winseq\), or of \(\spillseq\) when the top fragment is empty; in both cases it is
the last cell of the decoded stack.  Both cases of a depth-\(k\) access select
the cell at logical index \(|\spillseq|+|\winseq|-1-k\).  Depth returns
\(|\spillseq|+|\winseq|\), reset yields the empty stack, and a top-fragment policy
step preserves \(\spillseq\mathbin{\|}\winseq\) by definition.

The theorem covers a family rather than one proposal.  The fragmented
\layout{} of \cref{app:three-regions}, the flat ten-cell cache at
\(\nframe=0\), the fixed call-entry discipline, and an adaptive per-word entry
target all satisfy its hypotheses for successful transitions within the
capacity-reserved domain, which is why \cref{sec:rq3} treats the choice among
them as a performance question.

The theorem establishes agreement on stack contents and returned values for
successful modeled operations on one call-shared stack, and
\cref{app:two-stacks} extends this agreement to the data and floating-point
stacks together.  Exception snapshot and restoration, control-stack behaviour,
dictionary mutation, I/O, and the host compiler's guard recovery lie outside
the model, and the argument is on paper rather than mechanized.

\subsection{Encoding the top fragment in three regions}
\label{app:three-regions}

\Cref{thm:refinement} treats every cell outside the spill as one top fragment,
whereas the implementation splits that top fragment again into scalar fields and a
fixed array.  This section shows that the further split preserves the
meaning.

Let \(\mathit{frame}\) be the occupied prefix of a fixed array with
\(\nframe\) positions, ordered from bottom to top, and \(\mathit{top}\) the
occupied scalar fields in the same order, numbered from the top: if
\(\mathit{top}\) has \(m\) cells, exactly \(t_0,\ldots,t_{m-1}\) are
occupied and \(t_0\) holds the topmost value; \(\mathit{frame}=[a,b]\) and \(\mathit{top}=[c,d]\) mean the
top fragment is \([a,b,c,d]\) with \(t_0=d\) and \(t_1=c\).  A concrete state
\(c=\langle\spillseq,\mathit{frame},\mathit{top}\rangle\) is valid when
\(|\mathit{frame}|\le\nframe\), \(|\mathit{top}|\le\ntop\),
\(|\mathit{frame}|>0\Longrightarrow|\mathit{top}|=\ntop\), and
\(|\spillseq|+|\mathit{frame}|+|\mathit{top}|\le\dmax\); the implication states
the fill order, that scalar positions fill before the frame.  The top-fragment width
is \(\wmax=\ntop+\nframe\), and unused positions may hold stale values.  That
fill order determines a unique \emph{frame--top decomposition} of any top fragment
sequence \(W\) with \(|W|\le\wmax\): split
\(W=\mathit{frame}\mathbin{\|}\mathit{top}\) with
\(|\mathit{top}|=\min(|W|,\ntop)\).  Erasing the scalar--frame split gives
\(\project(c)=\langle\spillseq,\mathit{frame}\mathbin{\|}\mathit{top}\rangle\),
so the source-level stack represented by \(c\) is \(\decode(\project(c))\).
The concrete operations maintain this projection.  When \(\ntop>0\), push
fills the scalar positions first and then displaces the deepest scalar value
into the frame, and a full-fragment push first moves the deepest top-fragment cell to
the spill; a pop from a nonempty top fragment removes the top scalar value and, if the frame is nonempty,
moves its last value into the deepest scalar position.  When \(\ntop=0\), push
appends to the frame and a pop from a nonempty top fragment removes its last value, with the same spill rule for a
full-fragment push.  When the top fragment is empty, pop removes the last spill value,
and pop never refills the top fragment from the spill.  A
depth-\(k\) read or write selects \(t_k\), a frame position, or a spill position by
comparing \(k\) with \(|\mathit{top}|\) and \(|\mathit{top}|+|\mathit{frame}|\); and a top-fragment policy step of
\cref{app:refinement} acts on the two-region state and puts the resulting
top fragment into its frame--top decomposition.

\begin{restatable}[Three-region simulation]{lemma}{lemThreeRegion}
\label{lem:three-region}
Let \(c\) be a valid concrete state, and let a two-region operation be enabled
at \(\project(c)\) under the preconditions of \cref{app:operations,app:refinement}.
Then the corresponding concrete rule applies at \(c\) and yields a valid
\(c'\), the two-region operation produces \(\project(c')\), and any returned
values agree.
\end{restatable}

Proof sketch: write \(F=\mathit{frame}\) and \(T=\mathit{top}\).  For each
concrete case we compare the spill, \(F\mathbin{\|}T\), and any returned value
with the two-region rule.  A valid \(c\) projects to a valid two-region state,
so once the transitions agree, \cref{lem:invariant} gives the bounds on total
depth and spill length; in particular, a push enabled in the model raises the
total depth to at most \(\dmax\le\bspill\), leaving room to spill one cell when
the top fragment is full.  The cases below also preserve the individual region
bounds and the fill order.  The sequences contain only occupied positions, so
stale values in inactive slots do not affect the projection.
\begin{itemize}
\item \emph{Push with \(|T|<\ntop\).}  The fill order makes \(F\) empty, and
  \(x\) is appended to \(T\), so the top fragment becomes \(T\mathbin{\|}[x]\).
\item \emph{Push with \(|T|=\ntop>0\) and \(|F|<\nframe\).}  Write
  \(T=[y]\mathbin{\|}r\).  The deepest scalar value \(y\) moves to the end of
  the frame and \(x\) enters the scalar region, giving
  \(F\mathbin{\|}[y]\mathbin{\|}r\mathbin{\|}[x]=F\mathbin{\|}T\mathbin{\|}[x]\).
\item \emph{Push with a full top fragment and \(\ntop>0\).}  Write
  \(T=[y]\mathbin{\|}r\).  If \(\nframe>0\), write \(F=[z]\mathbin{\|}F'\);
  push moves \(z\) to the spill and produces
  \(\langle\spillseq\mathbin{\|}[z],F'\mathbin{\|}[y],r\mathbin{\|}[x]\rangle\).
  If \(\nframe=0\), push instead produces
  \(\langle\spillseq\mathbin{\|}[y],[],r\mathbin{\|}[x]\rangle\).  Both
  outcomes move the deepest top-fragment cell to the spill and append \(x\) to the
  top fragment, which is the full-fragment rule of \cref{app:operations}, and leave
  both regions at their configured widths.
\item \emph{Pop with \(F\) nonempty and \(\ntop>0\).}  The fill order makes \(T\) full.  Write
  \(T=r\mathbin{\|}[x]\) and \(F=F'\mathbin{\|}[z]\).  Pop returns \(x\), and
  the last frame value \(z\) moves into the deepest scalar position, giving
  \(F'\mathbin{\|}[z]\mathbin{\|}r=F\mathbin{\|}r\), which is the top fragment
  without its top cell.  \(T\) is full again, so the fill order holds.
\item \emph{Pop with \(F\) empty.}  Pop removes the last cell of \(T\), or the
  last spill cell when \(T\) is also empty, as in the two-region rule.  It is
  the only pop case that touches the spill, and it never moves spill cells back
  into the top fragment.
\item \emph{Indexed access at depth \(k\).}  The concrete rule selects
  \(t_k\) when \(k<|T|\), the frame position \(|F|-1-(k-|T|)\) when
  \(k<|T|+|F|\), and otherwise the spill position of the two-region rule.
  Because the top fragment is \(F\mathbin{\|}T\) with \(t_0\) on top, the first two
  cases select the top-fragment cell \(k\) positions below the top, so reads return,
  and writes replace, the same cell in both representations.
\item \emph{Reset, policy steps, depth.}  Reset empties all three sequences.
  A top-fragment policy step applies its two-region transition and then assigns the
  resulting top fragment to its frame--top decomposition, which satisfies the fill
  order by construction.  Depth is \(|\spillseq|+|F|+|T|\) in both
  representations.
\end{itemize}
With \(\ntop=0\), \(T=[]\) and \(\nframe=\wmax\ge1\).  If \(|F|<\nframe\),
push appends \(x\) to \(F\) and leaves the spill unchanged.  If
\(F=[z]\mathbin{\|}F'\) is full, push produces
\(\langle\spillseq\mathbin{\|}[z],F'\mathbin{\|}[x],[]\rangle\).  If
\(F=F'\mathbin{\|}[x]\), pop returns \(x\) and leaves
\(\langle\spillseq,F',[]\rangle\); if \(F=[]\), the empty-fragment pop case
above applies.  These are the two-region rules with \(F\) as the entire
top fragment, and the fill order holds with \(T=[]\).  With \(\nframe=0\), the
cases with a nonempty frame do not arise.

\begin{restatable}[Concrete stack refinement]{corollary}{corConcrete}
\label{cor:concrete-refinement}
Let \(c_0\) be a valid concrete state, and initialize an abstract Forth stack
to \(\decode(\project(c_0))\).  After any finite sequence of operations, each
enabled in the two-region model and matched by the corresponding abstract
operation, or by the identity for a top-fragment policy step, decoding the projection of the final concrete state
gives the abstract stack, and all returned values agree.
\end{restatable}

Proof sketch: induction on the operation sequence, composing
\cref{lem:three-region} with \cref{thm:refinement} at each step.

The evaluated default is \(\ntop=2\), \(\nframe=8\), \(\wmax=10\): scalar
fields make constant-depth operations direct, the fixed frame supplies eight
constant-index cells without one field per possible stack cell, and the spill
preserves depth to the configured capacity.  \Cref{fig:layout}
draws push and pop as before--after states.  Normalization differs from a call frame: it appends the
deepest top-fragment cells after the cells already in the spill, and returning does
not recover the pre-call arrangement.

\lemExitBound*

Proof sketch: by the assumptions, the top fragment contributes its occupancy and at
most \(\wmax\) cell values, the spill contributes its reference and index
rather than its cells, whose current values are already in memory, and each
exposed location contributes constant metadata.

\Cref{lem:exit-bound} bounds only state introduced by the data-stack representation; a
guard record can also contain program values.  In the evaluated case
\(\ccall=\ntop=2\), call normalization moves exactly \(|\winseq|-2\) cells
when the top fragment holds more than two cells.

\subparagraph*{Explicit materialization cost.}
\label{prop:necessity}
In the concrete representation, explicitly copying each of the
\(|\winseq|\) occupied top-fragment cells into the spill takes
\(\Theta(1+|\winseq|)\) word operations, including its fixed setup cost.
This cost does not follow from optimizer visibility alone: some values may
already be synchronized, and an observer may
need only depth, a subset of cells, or a deferred representation.  In
particular, Forth exception semantics does not require a contiguous snapshot of
all stack values.  Our implementation's fixed-width exception record is a
separate design choice, whose cost \cref{sec:rq4} measures.  Context
materialization and deoptimization provide related examples of reconstructing
optimized state~\cite{deutsch84,holzle92}, not a universal runtime lower bound
for every observer.

\subsection{Two call-shared stacks}
\label{app:two-stacks}

Forth's floating-point stack is a second call-shared sequence with the same
push, pop, indexed access, depth, and reset operations, and \system{} gives it
a top fragment of its own (\cref{sec:window-implementation}).  A machine state is a
pair \(\langle c_d,c_f\rangle\) of concrete states, one for the data stack and
one for the floating-point stack, each valid for its own parameters
\(\ntop\), \(\nframe\), \(\dmax\), and \(\bspill\); write
\(W_d\) and \(W_f\) for their top-fragment widths \(\wmax=\ntop+\nframe\).  Its meaning is the pair
\(\langle\decode(\project(c_d)),\decode(\project(c_f))\rangle\) of logical
stacks.  The two components share no storage: each has its own scalar fields,
frame, occupancy, spill array, and spill index.  A transition of one component
therefore leaves the other unchanged, both concretely and after decoding.

The corollary applies to successful modeled steps within the
capacity-reserved domain of each stack.  It composes for words whose stack
effects decompose into successful steps on one component each, provided each
step receives the same arguments in the two executions; for \code{F@}, this
means the load returns the same value.  \code{S>F} pops a cell from the data stack and pushes a
floating-point value, \code{F<} pops two floating-point values and pushes a
flag onto the data stack, and \code{F@} pops an address from the data stack
and pushes the floating-point value stored there; the conversion, comparison, or load supplies the pushed value and
touches neither stack.  An ordinary colon call applies a top-fragment policy step to
each component: in the build with both top fragments (\cref{sec:float-fragment}) it normalizes the data-stack top fragment and
leaves the floating-point top fragment unchanged, which is the identity step, and a
tail transfer leaves both unchanged.  For a colon-word execution token,
\code{CATCH} saves the floating-point depth, and on unwind \system{} pops
excess floating-point cells or pushes \(0.0\) for missing ones to restore
that depth; the corollary covers these individual stack operations.  Exception
bookkeeping and snapshot restoration, including the data-stack snapshot of
\cref{sec:window-implementation} and the snapshot taken for a primitive
execution token, lie outside its scope.

\begin{restatable}[Two-stack refinement]{corollary}{corTwoStack}
\label{cor:two-stack}
Let \(\langle c_d,c_f\rangle\) be a pair of valid concrete states with
disjoint representation storage, and
initialize two abstract Forth stacks to the decoded projections of \(c_d\) and
\(c_f\).  Run any finite sequence of successful steps, each a push, pop, indexed
access, depth query, reset, or top-fragment policy step on one component, and apply
to the matching abstract stack the abstract operation with the same arguments,
or the identity for a top-fragment policy step.  Arguments may be values returned by
earlier steps.  Then decoding the projections of the final components gives
the two abstract stacks, and all returned values agree.
\end{restatable}

Proof sketch: induction on the sequence; \cref{cor:concrete-refinement} covers
the component a step touches, and the other component is unchanged on both
sides.

\Cref{lem:exit-bound} applies to each component on its own.  If both top fragments
satisfy its assumptions, reconstructing the two stack representations at an
exit requires at most \(W_d+W_f\) top-fragment cell values plus \(O(1)\)
control values, with \(O(W_d+W_f)\) representation-derived metadata,
independent of both spill sizes and both logical depths.  The model does not
say which of these values an exit record actually carries.  In \system{} the
floating-point top fragment lives in a separate object outside the virtualizable
frame (\cref{sec:window-implementation}), and a trace that performs no floating-point stack access, including
through exception handling, needs no floating-point top-fragment values for stack
reconstruction.

\section{Supplementary Evaluation}
\label{app:supplementary-evaluation}

This appendix holds the parts of the evaluation that the body summarizes.

\subsection{Word-set coverage}
\label{app:wordsets-appendix}

\Cref{tab:wordsets} measures what \system{} implements of the standard word
sets, counting a word as implemented when the interpreter defines it with its
standard behaviour.  Core is complete, the extensions cover what the measured
workloads need, and the Block, Facility, Locals, and Tools sets are largely absent.
The benchmark ports also keep a few system-specific adaptations that
\cref{sec:threats} reports.

\begin{table}[h]
\centering
\small
\caption{Words of each standard word set implemented in \system{}, counted
against the word list of dpANS6 Appendix~F.  Core extension words are listed
separately, and other extension words are counted with their base set.}
\label{tab:wordsets}
\begin{tabular}{@{}lr@{ / }lr@{\qquad}lr@{ / }lr@{}}
\toprule
word set & \multicolumn{2}{c}{words} & \% & word set & \multicolumn{2}{c}{words} & \% \\
\midrule
Core          & 133 & 133 & 100 & Floating     & 62 & 72 & 86 \\
Core ext      &  39 &  46 &  85 & Search order & 13 & 14 & 93 \\
Double        &  15 &  22 &  68 & String       &  8 &  8 & 100 \\
Exception     &   4 &   4 & 100 & Tools        &  7 & 18 & 39 \\
File          &  20 &  25 &  80 & Facility     &  3 &  9 & 33 \\
Memory        &   3 &   3 & 100 & Locals       &  1 &  3 & 33 \\
\multicolumn{4}{c}{}                 & Block        &  3 & 14 & 21 \\
\midrule
all word sets & 311 & 371 &  84 & \multicolumn{4}{c}{} \\
\bottomrule
\end{tabular}
\end{table}

\subsection{RQ2: Trace metrics in detail}
\label{app:rq2-appendix}

This section compares the fragment layout with the flat ten-cell cache of
\cref{sec:rq2-traces}, using the measurements described there.

\begin{figure}[h]
\centering
\begin{tikzpicture}
\begin{axis}[
  width=0.94\linewidth,height=3.9cm,
  xmin=-0.52,xmax=2.52,ymode=log,log basis y=2,
  xtick={0,1,2},
  xticklabels={JIT compile time,operations per trace,guard-exit values},
  xticklabel style={font=\scriptsize,align=center},
  tick label style={font=\scriptsize},
  label style={font=\small},
  ylabel={\shortstack{flat ten-cell cache /\\top fragment}},
  legend style={at={(0.5,1.02)},anchor=south,legend columns=3,font=\scriptsize,draw=none},
  axis line style={black!60},ymajorgrids,grid style={draw=black!12},
  error bars/y dir=both,error bars/y explicit
]
\addplot[black!70,line width=0.6pt,no marks,forget plot] coordinates {(-0.52,1) (2.52,1)};
\addplot[only marks,mark=*,mark size=1.3pt,mark options={draw=plotBlue,fill=white,line width=0.45pt},forget plot] coordinates {(-0.30500000,1.85481521) (-0.29852941,1.11287539) (-0.29205882,0.88694639) (-0.28558824,1.10583658) (-0.27911765,1.11969550) (-0.27264706,0.72951168) (-0.26617647,1.17420004) (-0.25970588,1.27512924) (-0.25323529,0.92905405) (-0.24676471,1.03945885) (-0.24029412,1.04872390) (-0.23382353,1.19369895) (-0.22735294,2.29476332) (-0.22088235,1.05692927) (-0.21441176,0.83458311) (-0.20794118,1.02530459) (-0.20147059,1.03937947) (-0.19500000,1.05049732) (0.69500000,0.67876461) (0.70147059,1.32524272) (0.70794118,1.17469880) (0.71441176,1.06574761) (0.72088235,1.16852540) (0.72735294,1.19749573) (0.73382353,1.31576977) (0.74029412,1.00174520) (0.74676471,1.15178571) (0.75323529,1.27800830) (0.75970588,1.29285714) (0.76617647,1.06363636) (0.77264706,0.87798783) (0.77911765,1.30731707) (0.78558824,1.13381942) (0.79205882,1.21374046) (0.79852941,1.34090909) (0.80500000,1.31878788) (1.69500000,1.02222222) (1.70147059,1.00000000) (1.70794118,1.04545455) (1.71441176,0.95833333) (1.72088235,1.00000000) (1.72735294,1.00000000) (1.73382353,1.00000000) (1.74029412,1.00000000) (1.74676471,1.00000000) (1.75323529,1.04166667) (1.75970588,0.97826087) (1.76617647,1.04545455) (1.77264706,0.83333333) (1.77911765,1.00000000) (1.78558824,1.04545455) (1.79205882,1.04761905) (1.79852941,1.05000000) (1.80500000,1.04761905)};
\addplot[only marks,mark=*,mark size=3.2pt,mark options={draw=black!80,fill=plotBlue,line width=0.6pt},error bars/error bar style={draw=plotBlue,line width=0.9pt}] table[x=x,y=y,y error minus=minus,y error plus=plus] {
x y minus plus
-0.25000000 1.11222157 0.09954112 0.12238332
0.75000000 1.14665756 0.07793873 0.06938793
1.75000000 1.00507541 0.02192353 0.01814621
};
\addlegendentry{Shootout kernels}
\addplot[only marks,mark=square*,mark size=1.3pt,mark options={draw=plotGreen,fill=white,line width=0.45pt},forget plot] coordinates {(-0.05500000,1.19082958) (-0.04653846,1.29347655) (-0.03807692,1.13839736) (-0.02961538,1.31234955) (-0.02115385,1.81995522) (-0.01269231,1.29648428) (-0.00423077,0.93978598) (0.00423077,0.81718268) (0.01269231,0.84385002) (0.02115385,0.52520661) (0.02961538,1.34860827) (0.03807692,0.15584416) (0.04653846,1.76868255) (0.05500000,2.10689094) (0.94500000,1.43288628) (0.95346154,1.03719665) (0.96192308,1.01149820) (0.97038462,0.98814117) (0.97884615,1.31038356) (0.98730769,0.93868528) (0.99576923,1.12877595) (1.00423077,0.66662667) (1.01269231,1.10648050) (1.02115385,1.16220763) (1.02961538,0.99284366) (1.03807692,1.48108589) (1.04653846,1.02006006) (1.05500000,0.89802480) (1.94500000,0.91304348) (1.95346154,0.90476190) (1.96192308,0.86363636) (1.97038462,0.86363636) (1.97884615,0.95238095) (1.98730769,0.95238095) (1.99576923,0.95454545) (2.00423077,0.95652174) (2.01269231,0.90909091) (2.02115385,1.00000000) (2.02961538,0.95454545) (2.03807692,1.00000000) (2.04653846,0.90476190) (2.05500000,0.86956522)};
\addplot[only marks,mark=square*,mark size=3.2pt,mark options={draw=black!80,fill=plotGreen,line width=0.6pt},error bars/error bar style={draw=plotGreen,line width=0.9pt}] table[x=x,y=y,y error minus=minus,y error plus=plus] {
x y minus plus
0.00000000 1.02676443 0.26734193 0.29401893
1.00000000 1.06394500 0.08876618 0.09126037
2.00000000 0.92744600 0.01908766 0.01933522
};
\addlegendentry{AWFY benchmarks}
\addplot[only marks,mark=triangle*,mark size=1.3pt,mark options={draw=plotOrange,fill=white,line width=0.45pt},forget plot] coordinates {(0.19500000,0.54013123) (0.21700000,0.77773708) (0.23900000,1.57334846) (0.26100000,0.71672352) (0.28300000,1.42403042) (0.30500000,1.34547539) (1.19500000,1.10768176) (1.21700000,1.45569703) (1.23900000,1.48908292) (1.26100000,1.27874235) (1.28300000,1.44165525) (1.30500000,1.30470978) (2.19500000,0.95238095) (2.21700000,0.90909091) (2.23900000,0.90909091) (2.26100000,1.00000000) (2.28300000,0.90909091) (2.30500000,0.95238095)};
\addplot[only marks,mark=triangle*,mark size=3.2pt,mark options={draw=black!80,fill=plotOrange,line width=0.6pt},error bars/error bar style={draw=plotOrange,line width=0.9pt}] table[x=x,y=y,y error minus=minus,y error plus=plus] {
x y minus plus
0.25000000 0.98397474 0.23955948 0.30440943
1.25000000 1.33945217 0.09259468 0.09070227
2.25000000 0.93808147 0.02191468 0.02280237
};
\addlegendentry{Appbench applications}
\end{axis}
\end{tikzpicture}
\caption{Ratios of the flat ten-cell cache to the
top-fragment layout on Zen~3.  Values above one mean less compilation time,
fewer operations per optimized trace, or fewer guard-exit values with the
top fragment.  Small marks are programs and large marks are suite geometric
means with 90\% bootstrap intervals.
}
\label{fig:jit-metrics}
\end{figure}

The clearest difference is trace length (\cref{fig:jit-metrics}).  On Zen~3,
the top-fragment layout emits about 13\% fewer operations per optimized trace on
Shootout, 6\% fewer on AWFY, and 25\% fewer on Appbench, with 90\% intervals of
1.07--1.22, 0.98--1.16, and 1.25--1.43 on the cache-to-fragment ratio.  It is
below the flat ten-cell cache on 16 of the 18 kernels, 9 of the 14 AWFY
benchmarks, and all six applications; on AWFY, \code{NBody} is the clear
exception (0.67).  Guard-exit records do not shrink with them.  They are nearly
unchanged on Shootout and about 8\% and 7\% larger on AWFY and Appbench.
Compilation time falls by about 10\% on Shootout (interval 1.01--1.24), but
\code{recurse} and \code{ack}, which compile 2.3\(\times\) and 1.9\(\times\)
faster, contribute much of it, and the median kernel gains 5\%.  The AWFY and
Appbench intervals, 0.76--1.32 and 0.74--1.29, include one, so we claim no
compile-time gain there.

Shorter traces do not come with fewer spill accesses or, on AWFY and Appbench,
with shorter guard-exit records (\cref{fig:jit-mechanism}).  Against the
flat ten-cell cache, both Shootout medians stay at zero residual spill
accesses per optimized trace, the AWFY median rises from about 1.78 to 7.10,
and the Appbench median from 0.85 to 3.67.  The median guard-exit record goes
from 23 to 22.25 values on Shootout and from 20 to 22 on AWFY and Appbench.  The
top fragment emits fewer operations in spite of those extra accesses, and these
counts alone do not say which transformations produce the timing differences.
Guard-exit values also include non-stack values, so they neither prove the
\(\wmax\) bound nor isolate its data-stack part, and the experiment holds the
configured capacity fixed rather than testing independence from logical
depth.

Sapphire Rapids gives similar trace metrics.  Its operations-per-trace
ratios are 1.16, 1.08, and 1.29 on Shootout, AWFY, and Appbench, against 1.15,
1.06, and 1.34 on Zen~3.  Its Shootout compile-time ratio is 1.18, again
mostly from \code{recurse} and \code{ack}, and its AWFY and Appbench intervals
include one.  The guard-exit and spill-access medians are within 0.25 values
of those on Zen~3.

\begin{figure}[tb]
\centering
\begin{tikzpicture}
\begin{groupplot}[
  group style={group size=2 by 1,horizontal sep=1.3cm},
  width=0.43\linewidth,height=3.3cm,
  xmin=-0.42,xmax=2.42,xtick={0,1,2},
  xticklabels={Shootout,AWFY,Appbench},
  tick label style={font=\scriptsize},
  label style={font=\scriptsize},
  axis line style={black!60},ymajorgrids,grid style={draw=black!12},
  legend style={at={(0,1.04)},anchor=south west,legend columns=2,font=\scriptsize,draw=none}
]
\nextgroupplot[ylabel={residual spill accesses / trace}]
\addplot[only marks,mark=square*,mark size=1.3pt,mark options={draw=plotOrange,fill=white,line width=0.45pt},forget plot] coordinates {(-0.16500000,1.72727273) (-0.15852941,0.00000000) (-0.15205882,0.00000000) (-0.14558824,0.00000000) (-0.13911765,0.00000000) (-0.13264706,0.00000000) (-0.12617647,0.00000000) (-0.11970588,0.09090909) (-0.11323529,0.00000000) (-0.10676471,0.00000000) (-0.10029412,0.00000000) (-0.09382353,0.00000000) (-0.08735294,2.96363636) (-0.08088235,0.00000000) (-0.07441176,0.00000000) (-0.06794118,0.00000000) (-0.06147059,0.00000000) (-0.05500000,0.00000000) (0.83500000,0.00000000) (0.84346154,1.43384982) (0.85192308,4.01333333) (0.86038462,2.19330855) (0.86884615,0.43750000) (0.87730769,2.57142857) (0.88576923,0.00000000) (0.89423077,0.00000000) (0.90269231,2.12903226) (0.91115385,2.76744186) (0.91961538,0.40350877) (0.92807692,0.00000000) (0.93653846,2.55555556) (0.94500000,17.39285714) (1.83500000,1.38095238) (1.85700000,0.32792208) (1.87900000,0.00000000) (1.90100000,1.89965398) (1.92300000,0.28531469) (1.94500000,1.58426966)};
\addplot[only marks,mark=square*,mark size=3.2pt,mark options={draw=black!80,fill=plotOrange,line width=0.6pt}] coordinates {(-0.11000000,0.00000000) (0.89000000,1.78144104) (1.89000000,0.85443723)};
\addlegendentry{10-cell cache}
\addplot[only marks,mark=*,mark size=1.3pt,mark options={draw=plotBlue,fill=white,line width=0.45pt},forget plot] coordinates {(0.05500000,6.33333333) (0.06147059,0.00000000) (0.06794118,0.00000000) (0.07441176,0.00000000) (0.08088235,0.00000000) (0.08735294,0.00000000) (0.09382353,0.00000000) (0.10029412,3.36363636) (0.10676471,0.00000000) (0.11323529,0.00000000) (0.11970588,0.00000000) (0.12617647,0.00000000) (0.13264706,7.00000000) (0.13911765,0.00000000) (0.14558824,0.00000000) (0.15205882,0.00000000) (0.15852941,0.00000000) (0.16500000,0.00000000) (1.05500000,4.80769231) (1.06346154,7.38896189) (1.07192308,15.30147059) (1.08038462,13.13247863) (1.08884615,2.84112150) (1.09730769,7.96428571) (1.10576923,0.00000000) (1.11423077,8.72727273) (1.12269231,4.47826087) (1.13115385,3.17021277) (1.13961538,15.34615385) (1.14807692,0.00000000) (1.15653846,6.81818182) (1.16500000,38.93750000) (2.05500000,3.52100840) (2.07700000,1.28672316) (2.09900000,8.68102288) (2.12100000,3.81424149) (2.14300000,0.72760018) (2.16500000,6.25510204)};
\addplot[only marks,mark=*,mark size=3.2pt,mark options={draw=black!80,fill=plotBlue,line width=0.6pt}] coordinates {(0.11000000,0.00000000) (1.11000000,7.10357186) (2.11000000,3.66762494)};
\addlegendentry{top fragment}
\nextgroupplot[ylabel={values in a guard-exit record}]
\addplot[only marks,mark=square*,mark size=1.3pt,mark options={draw=plotOrange,fill=white,line width=0.45pt},forget plot] coordinates {(-0.16500000,23.00000000) (-0.15852941,22.00000000) (-0.15205882,23.00000000) (-0.14558824,23.00000000) (-0.13911765,24.00000000) (-0.13264706,22.00000000) (-0.12617647,23.00000000) (-0.11970588,23.00000000) (-0.11323529,23.00000000) (-0.10676471,25.00000000) (-0.10029412,22.50000000) (-0.09382353,23.00000000) (-0.08735294,20.00000000) (-0.08088235,21.00000000) (-0.07441176,23.00000000) (-0.06794118,22.00000000) (-0.06147059,21.00000000) (-0.05500000,22.00000000) (0.83500000,21.00000000) (0.84346154,19.00000000) (0.85192308,19.00000000) (0.86038462,19.00000000) (0.86884615,20.00000000) (0.87730769,20.00000000) (0.88576923,21.00000000) (0.89423077,22.00000000) (0.90269231,20.00000000) (0.91115385,21.00000000) (0.91961538,21.00000000) (0.92807692,22.00000000) (0.93653846,19.00000000) (0.94500000,20.00000000) (1.83500000,20.00000000) (1.85700000,20.00000000) (1.87900000,20.00000000) (1.90100000,22.00000000) (1.92300000,20.00000000) (1.94500000,20.00000000)};
\addplot[only marks,mark=square*,mark size=3.2pt,mark options={draw=black!80,fill=plotOrange,line width=0.6pt}] coordinates {(-0.11000000,23.00000000) (0.89000000,20.00000000) (1.89000000,20.00000000)};
\addplot[only marks,mark=*,mark size=1.3pt,mark options={draw=plotBlue,fill=white,line width=0.45pt},forget plot] coordinates {(0.05500000,22.50000000) (0.06147059,22.00000000) (0.06794118,22.00000000) (0.07441176,24.00000000) (0.08088235,24.00000000) (0.08735294,22.00000000) (0.09382353,23.00000000) (0.10029412,23.00000000) (0.10676471,23.00000000) (0.11323529,24.00000000) (0.11970588,23.00000000) (0.12617647,22.00000000) (0.13264706,24.00000000) (0.13911765,21.00000000) (0.14558824,22.00000000) (0.15205882,21.00000000) (0.15852941,20.00000000) (0.16500000,21.00000000) (1.05500000,23.00000000) (1.06346154,21.00000000) (1.07192308,22.00000000) (1.08038462,22.00000000) (1.08884615,21.00000000) (1.09730769,21.00000000) (1.10576923,22.00000000) (1.11423077,23.00000000) (1.12269231,22.00000000) (1.13115385,21.00000000) (1.13961538,22.00000000) (1.14807692,22.00000000) (1.15653846,21.00000000) (1.16500000,23.00000000) (2.05500000,21.00000000) (2.07700000,22.00000000) (2.09900000,22.00000000) (2.12100000,22.00000000) (2.14300000,22.00000000) (2.16500000,21.00000000)};
\addplot[only marks,mark=*,mark size=3.2pt,mark options={draw=black!80,fill=plotBlue,line width=0.6pt}] coordinates {(0.11000000,22.25000000) (1.11000000,22.00000000) (2.11000000,22.00000000)};
\end{groupplot}
\end{tikzpicture}
\caption{Direct mechanism measurements from optimized loops and bridges on
Zen~3.  Spill accesses are residual array operations whose base value was
loaded from the interpreter's shared-spill field; guard-exit values count every
value in the record, not only data-stack cells.  Small marks are programs, large
marks suite medians.}
\label{fig:jit-mechanism}
\end{figure}

\subsection{RQ3: Drift checks and further shapes}
\label{app:rq3-appendix}

\subparagraph*{Measurement drift.}
All 228 program--configuration cells on each machine completed, the Shootout
and AWFY cells with their expected output, and the primary analysis keeps the
cells that drift.  On both machines some cells exceeded the prespecified 8\%
difference between the medians of the two halves of the fixed final-half
region.  Two checks test the speedup of the complete layout over the two-cell
cache (Zen~3/Sapphire Rapids).  Dropping every program in which any of the
four main configurations crossed the threshold gives 1.05\(\times\)/1.08\(\times\) on
Shootout, 1.25\(\times\)/1.18\(\times\) on AWFY, and
1.28\(\times\)/1.18\(\times\) on Appbench.  Leave-one-program-out estimates
over all cells span 1.04--1.10 and 1.07--1.15 on Shootout, 1.22--1.28 and
1.14--1.20 on AWFY, and 1.16--1.28 and 1.11--1.18 on Appbench.  The point
estimates stay above one in both checks.  This says nothing about every
pairwise policy contrast.

\subparagraph*{Other shapes and adaptive entry.}
With ten exposed cells, a four-scalar/six-frame split improves the Appbench median
by about 4--6\%, whereas eight scalar positions lose on both suites and
machines.  Holding the scalars at two, frame widths from four to 32 stay near
the default for most programs, and the clearest loss at large widths is the
exception-heavy kernel of \cref{sec:rq4}.  Fragmented adaptive entry raises the
Appbench median only on Zen~3, from 1.19 to 1.21 over the two-cell cache,
widens the interquartile range on both machines, and regresses on
\code{ack}, so it does not dominate the fixed policy.  Over the sixteen
non-reference suite--setting pairs, the two machines' median speedups differ
by 0.6\% at the median.

\subsection{RQ4: Where the copying is paid}
\label{app:rq4-appendix}

The three copying operations of \cref{sec:rq4} are measured in turn.

\subparagraph*{Exception snapshots.}
The exception-heavy kernel \code{except} slows monotonically as the top fragment
widens.  It reaches 0.74\(\times\), 0.54\(\times\), and 0.33\(\times\) of the
default two-scalar/eight-frame configuration at \(\nframe=16\), 32, and 64 on
Zen~3 and 0.78\(\times\), 0.55\(\times\), and 0.35\(\times\) on Sapphire
Rapids.  At \(\nframe=64\) the next-lowest kernel is at 0.84\(\times\) and
0.94\(\times\).  The kernel enters and leaves catch
frames on its hot path, and a frame saves every configured top-fragment position, so
a wider top fragment means a larger snapshot even when the occupancy does not
change.  This points to the fixed-width copying cost of \cref{sec:integration} rather
than a cost inherent to exception handling, but the sweep does not isolate it,
since width also changes what a full-fragment push shifts.

To separate the snapshot itself we compare fixed-slot with occupancy-aware
copying at two frame widths, leaving layout and call-entry policy alone.  Every
\code{CATCH} in this workload snapshots an empty top fragment, because the execution
token has been consumed and the nested words take no data arguments.  The
opt-in alternative skips inactive slots, so here it copies no frame slot.

\begin{figure}[t]
\centering
\begin{tikzpicture}
\begin{axis}[xmin=0.8,xmax=4.9,ymin=-0.6,ymax=3.6,ytick={0,1,2,3},yticklabels={Zen 3 width 8,Zen 3 width 64,Sapphire Rapids width 8,Sapphire Rapids width 64},tick label style={font=\scriptsize},label style={font=\small},axis line style={black!60},xmajorgrids,grid style={draw=black!12},xlabel={fixed-slot / live-frame time},width=0.78\linewidth,height=3.0cm]
\addplot[only marks,mark=*,mark size=2.1pt,mark options={draw=plotBlue,fill=plotBlue,line width=0.7pt},error bars/.cd,x dir=both,x explicit,error bar style={draw=plotBlue,line width=0.9pt}] table[x=value,y=y,x error minus=minus,x error plus=plus] {
value y minus plus
1.53703704 0 0.07407407 0.11111111
};
\addplot[only marks,mark=*,mark size=2.1pt,mark options={draw=plotBlue,fill=plotBlue,line width=0.7pt},error bars/.cd,x dir=both,x explicit,error bar style={draw=plotBlue,line width=0.9pt}] table[x=value,y=y,x error minus=minus,x error plus=plus] {
value y minus plus
4.40740741 1 0.09831650 0.01851852
};
\addplot[only marks,mark=*,mark size=2.1pt,mark options={draw=plotBlue,fill=white,line width=0.7pt},error bars/.cd,x dir=both,x explicit,error bar style={draw=plotBlue,line width=0.9pt}] table[x=value,y=y,x error minus=minus,x error plus=plus] {
value y minus plus
1.58333333 2 0.00000000 0.00000000
};
\addplot[only marks,mark=*,mark size=2.1pt,mark options={draw=plotBlue,fill=white,line width=0.7pt},error bars/.cd,x dir=both,x explicit,error bar style={draw=plotBlue,line width=0.9pt}] table[x=value,y=y,x error minus=minus,x error plus=plus] {
value y minus plus
4.45901639 3 0.01639344 0.00000000
};
\end{axis}
\end{tikzpicture}
\caption{Runtime of \code{except} with fixed-slot snapshots divided by its
runtime with occupancy-aware snapshots.  Points show the single-kernel estimate at each frame width;
bars are 90\% bootstrap intervals over 100 fresh processes.  Frame width
excludes the two scalar top fields.}
\label{fig:live-snapshot}
\end{figure}

Much of the fixed-slot work is therefore avoidable: the speedup grows from
\(1.54\times\) at eight frame slots to \(4.41\times\) at 64 on Zen~3 and from
\(1.58\times\) to \(4.46\times\) on Sapphire Rapids (\cref{fig:live-snapshot}).
The Sapphire Rapids interval at eight slots has zero width at the recorded
timing resolution, which does not make the estimate certain.  These are empty snapshots, so the figure is neither a curve over
partial occupancy nor a claim about other programs.

\subparagraph*{Normalization.}
The second price grows with the occupancy at a call boundary, and on Forth
that is low: the median program moves an average of 0.04 cells per normalization on
Shootout, 0.41 on AWFY, and 0.60 on Appbench.  On Appbench, normalization changes the spread rather than the suite
estimate (speedups over the two-cell cache, Zen~3/Sapphire Rapids), lifting
\code{fcp} from 0.849 to 1.121 and from 0.882 to 1.111 while lowering
\code{lexex} from 1.379 to 1.135 and from 1.241 to 1.027 (\cref{sec:rq3}).  In
RPyFactor normalization may explain the losses on \code{heaps}, \code{fannkuch}, and \code{euler150}, whose
inner loops keep cells below the entry target and read them again
(\cref{sec:factor}).

\subparagraph*{Materialization.}
The third price is materialization.  Forcing one at depths 2, 8, 32, and 128
gives time ratios over the same operation without it of 1.17, 0.71, 0.96, and 1.01 on Zen~3 and 1.11, 0.65,
0.96, and 0.99 on Sapphire Rapids.  Above the top-fragment width the point estimates stay within 4\% of
one, consistent with the depth-independent \(O(\wmax)\) bound of
\cref{tab:costs}.  These are whole-operation ratios rather than copy latencies,
and the Zen~3 interval at depth 128 reaches 1.14.

\section{Top-Fragment Ablation Table}
\label{app:ablation-table}

\begin{table}[H]
\centering
\small
\caption{Time of each reduced build divided by the time of the build with both
top fragments (Zen~3), with 90\% bootstrap intervals; the AWFY rows split the suite
into its three floating-point and eleven integer-only programs.}
\label{tab:frag-ablation}
\resizebox{0.8\linewidth}{!}{\begin{tabular}{lrrr}
\toprule
Suite & no float fragment & no integer fragment & neither fragment \\
\midrule
Shootout & 1.01 [0.98, 1.03] & 1.83 [1.42, 2.39] & 1.84 [1.43, 2.42] \\
AWFY & 1.20 [1.03, 1.45] & 2.42 [2.11, 2.74] & 2.73 [2.45, 3.05] \\
AWFY (float) & 2.31 [1.63, 3.27] & 1.86 [1.42, 2.43] & 3.37 [2.85, 3.98] \\
AWFY (int) & 1.01 [1.00, 1.02] & 2.60 [2.31, 2.92] & 2.58 [2.31, 2.90] \\
Appbench & 0.99 [0.96, 1.01] & 2.46 [1.96, 3.05] & 2.50 [2.04, 3.05] \\
\bottomrule
\end{tabular}
}
\end{table}

\section{Dynamic Workload Characteristics}
\label{app:workloads}

\begin{figure}[H]
\centering
\begin{tikzpicture}
\begin{groupplot}[
  group style={group size=2 by 2,horizontal sep=1.2cm,vertical sep=1.35cm},
  width=0.43\linewidth,height=3.1cm,
  xmin=-0.35,xmax=2.35,xtick={0,1,2},
  xticklabels={Shootout,AWFY,Appbench},
  tick label style={font=\scriptsize},label style={font=\scriptsize},
  axis line style={black!60},ymajorgrids,grid style={draw=black!12},
  title style={yshift=-.6ex}
]
\nextgroupplot[title={integer-stack 95th percentile},title style={font=\scriptsize}]
\addplot[only marks,mark=*,mark size=1.4pt,mark options={draw=plotOrange,fill=white,line width=0.45pt},forget plot] coordinates {(-0.0720,31.00000000) (-0.0540,4.00000000) (-0.0360,2.00000000) (-0.0180,3.00000000) (0.0000,4.00000000) (0.0180,4.00000000) (0.0360,3.00000000) (0.0540,13.00000000) (0.0720,2.00000000) (-0.0720,2.00000000) (-0.0540,3.00000000) (-0.0360,3.00000000) (-0.0180,31.00000000) (0.0000,31.00000000) (0.0180,5.00000000) (0.0360,5.00000000) (0.0540,5.00000000) (0.0720,3.00000000)};
\addplot[only marks,mark=*,mark size=3.2pt,mark options={draw=black!80,fill=plotOrange,line width=0.6pt},forget plot] coordinates {(0,4.00000000)};
\addplot[only marks,mark=*,mark size=1.4pt,mark options={draw=plotGreen,fill=white,line width=0.45pt},forget plot] coordinates {(0.9280,9.00000000) (0.9460,22.00000000) (0.9640,21.00000000) (0.9820,31.00000000) (1.0000,16.00000000) (1.0180,31.00000000) (1.0360,8.00000000) (1.0540,9.00000000) (1.0720,20.00000000) (0.9280,22.00000000) (0.9460,8.00000000) (0.9640,10.00000000) (0.9820,25.00000000) (1.0000,31.00000000)};
\addplot[only marks,mark=*,mark size=3.2pt,mark options={draw=black!80,fill=plotGreen,line width=0.6pt},forget plot] coordinates {(1,20.50000000)};
\addplot[only marks,mark=*,mark size=1.4pt,mark options={draw=plotBlue,fill=white,line width=0.45pt},forget plot] coordinates {(1.9280,10.00000000) (1.9460,19.00000000) (1.9640,6.00000000) (1.9820,21.00000000) (2.0000,19.00000000) (2.0180,10.00000000)};
\addplot[only marks,mark=*,mark size=3.2pt,mark options={draw=black!80,fill=plotBlue,line width=0.6pt},forget plot] coordinates {(2,14.50000000)};
\nextgroupplot[title={word calls / 1,000 primitives},title style={font=\scriptsize},ymode=log,log basis y=10]
\addplot[only marks,mark=*,mark size=1.4pt,mark options={draw=plotOrange,fill=white,line width=0.45pt},forget plot] coordinates {(-0.0720,47.61398884) (-0.0540,0.00001563) (-0.0360,44.11762933) (-0.0180,79.99988587) (0.0000,0.00539139) (0.0180,55.83045771) (0.0360,0.00000137) (0.0540,0.00049380) (0.0720,0.00000097) (-0.0720,0.00000064) (-0.0540,0.00014285) (-0.0360,0.00000175) (-0.0180,67.79660044) (0.0000,0.01388445) (0.0180,0.00279676) (0.0360,38.45795927) (0.0540,0.00058823) (0.0720,0.01176060)};
\addplot[only marks,mark=*,mark size=3.2pt,mark options={draw=black!80,fill=plotOrange,line width=0.6pt},forget plot] coordinates {(0,0.00409407)};
\addplot[only marks,mark=*,mark size=1.4pt,mark options={draw=plotGreen,fill=white,line width=0.45pt},forget plot] coordinates {(0.9280,72.48220043) (0.9460,42.74822229) (0.9640,80.87633519) (0.9820,64.47940492) (1.0000,31.27951524) (1.0180,7.85979351) (1.0360,0.00000769) (1.0540,13.27057591) (1.0720,33.50313130) (0.9280,22.67485493) (0.9460,55.99641351) (0.9640,0.00687461) (0.9820,18.69293384) (1.0000,55.63066730)};
\addplot[only marks,mark=*,mark size=3.2pt,mark options={draw=black!80,fill=plotGreen,line width=0.6pt},forget plot] coordinates {(1,32.39132327)};
\addplot[only marks,mark=*,mark size=1.4pt,mark options={draw=plotBlue,fill=white,line width=0.45pt},forget plot] coordinates {(1.9280,14.80684407) (1.9460,30.53312510) (1.9640,155.33538686) (1.9820,3.76012361) (2.0000,39.28077031) (2.0180,6.98020211)};
\addplot[only marks,mark=*,mark size=3.2pt,mark options={draw=black!80,fill=plotBlue,line width=0.6pt},forget plot] coordinates {(2,22.66998459)};
\nextgroupplot[title={spill accesses (\%)},title style={font=\scriptsize},ymode=log,log basis y=10]
\addplot[only marks,mark=*,mark size=1.4pt,mark options={draw=plotOrange,fill=white,line width=0.45pt},forget plot] coordinates {(-0.0720,8.17297801) (-0.0540,0.00000040) (-0.0360,0.00000357) (-0.0180,0.00000860) (0.0000,0.00128634) (0.0180,0.00121538) (0.0360,0.00000011) (0.0540,0.00009032) (0.0720,0.00000027) (-0.0720,0.00000026) (-0.0540,0.00002500) (-0.0360,0.00000029) (-0.0180,27.31889057) (0.0000,1.06477444) (0.0180,0.00020562) (0.0360,0.00016663) (0.0540,0.00005263) (0.0720,0.00000417)};
\addplot[only marks,mark=*,mark size=3.2pt,mark options={draw=black!80,fill=plotOrange,line width=0.6pt},forget plot] coordinates {(0,0.00003882)};
\addplot[only marks,mark=*,mark size=1.4pt,mark options={draw=plotGreen,fill=white,line width=0.45pt},forget plot] coordinates {(0.9280,9.13959000) (0.9460,7.75637950) (0.9640,5.20368686) (0.9820,8.67850883) (1.0000,1.02502836) (1.0180,2.42196762) (1.0360,0.00000183) (1.0540,4.83714524) (1.0720,3.91989359) (0.9280,8.51333467) (0.9460,9.76384282) (0.9640,0.00000012) (0.9820,4.71599920) (1.0000,17.56822867)};
\addplot[only marks,mark=*,mark size=3.2pt,mark options={draw=black!80,fill=plotGreen,line width=0.6pt},forget plot] coordinates {(1,5.02041605)};
\addplot[only marks,mark=*,mark size=1.4pt,mark options={draw=plotBlue,fill=white,line width=0.45pt},forget plot] coordinates {(1.9280,2.36364477) (1.9460,1.25855394) (1.9640,2.33205921) (1.9820,1.60741599) (2.0000,0.27504853) (2.0180,3.26740438)};
\addplot[only marks,mark=*,mark size=3.2pt,mark options={draw=black!80,fill=plotBlue,line width=0.6pt},forget plot] coordinates {(2,1.96973760)};
\nextgroupplot[title={cells moved / normalization},title style={font=\scriptsize},ymode=log,log basis y=10]
\addplot[only marks,mark=*,mark size=1.4pt,mark options={draw=plotOrange,fill=white,line width=0.45pt},forget plot] coordinates {(-0.0720,1.99405495) (-0.0540,0.02631579) (-0.0360,0.00000100) (-0.0180,0.00000033) (0.0000,0.04651163) (0.0180,0.00016595) (0.0360,0.03636364) (0.0540,0.15686275) (0.0720,0.04347826) (-0.0720,0.04545455) (-0.0540,0.04255319) (-0.0360,0.03773585) (-0.0180,1.17467107) (0.0000,0.15686275) (0.0180,0.04000000) (0.0360,0.00003995) (0.0540,0.04255319) (0.0720,0.00246609)};
\addplot[only marks,mark=*,mark size=3.2pt,mark options={draw=black!80,fill=plotOrange,line width=0.6pt},forget plot] coordinates {(0,0.04127660)};
\addplot[only marks,mark=*,mark size=1.4pt,mark options={draw=plotGreen,fill=white,line width=0.45pt},forget plot] coordinates {(0.9280,0.34174918) (0.9460,0.41928113) (0.9640,0.19241421) (0.9820,0.57844650) (1.0000,0.27027917) (1.0180,1.49248982) (1.0360,0.40000000) (1.0540,0.16129061) (1.0720,0.54118238) (0.9280,0.22861490) (0.9460,0.53577042) (0.9640,0.00033300) (0.9820,0.74990841) (1.0000,0.93768233)};
\addplot[only marks,mark=*,mark size=3.2pt,mark options={draw=black!80,fill=plotGreen,line width=0.6pt},forget plot] coordinates {(1,0.40964056)};
\addplot[only marks,mark=*,mark size=1.4pt,mark options={draw=plotBlue,fill=white,line width=0.45pt},forget plot] coordinates {(1.9280,0.94420493) (1.9460,0.45906008) (1.9640,0.13182178) (1.9820,0.82692513) (2.0000,0.02714506) (2.0180,0.74720982)};
\addplot[only marks,mark=*,mark size=3.2pt,mark options={draw=black!80,fill=plotBlue,line width=0.6pt},forget plot] coordinates {(2,0.60313495)};
\end{groupplot}
\end{tikzpicture}
\caption{Dynamic characteristics of the programs used in the Forth suite
comparisons.  Small marks are programs, large marks suite medians.  Event
counts come from an instrumented execution and are not timing samples; the
depth counter pools depths of 31 or more in its final bin.}
\label{fig:workloads}
\end{figure}

\end{document}